\documentclass[aps,prb, reprint,superscriptaddress]{revtex4-1}
\usepackage{amssymb}
\usepackage{amsmath}
\usepackage{graphicx} 
\usepackage[section]{placeins}

\begin{document}

\preprint{1}

\title{Influence of the FFLO-like State on the Upper Critical Field of a S/F Bilayer: Angular and Temperature Dependence} 



\author{D.~Lenk}
\affiliation{Experimentalphysik II, Institut f\"{u}r Physik, Universit\"{a}t Augsburg, Universit\"{a}tsstra{\ss}e 1, D-86159 Augsburg, Germany}

\author{M.~Hemmida}
\affiliation{Experimentalphysik II, Institut f\"{u}r Physik, Universit\"{a}t Augsburg, Universit\"{a}tsstra{\ss}e 1, D-86159 Augsburg, Germany}
\affiliation{Experimentalphysik V, Center for Electronic Correlations and Magnetism, Institut 
f\"{u}r Physik, Universit\"{a}t Augsburg, Universit\"{a}tsstra{\ss}e 1, D-86159 Augsburg, Germany}

\author{R. Morari}
\affiliation{Experimentalphysik II, Institut f\"{u}r Physik, Universit\"{a}t Augsburg, Universit\"{a}tsstra{\ss}e 1, D-86159 Augsburg, Germany}
\affiliation{D.~Ghitsu Institute of Electronic Engineering and Nanotechnologies ASM,Academiei Str. 3/3, MD2028 Kishinev, Moldova}
\affiliation{Solid State Physics Department, Kazan Federal University, Kremlevskaya Str. 18, 420008 Kazan, Russian Federation}

\author{V. I.~Zdravkov}
\affiliation{Experimentalphysik II, Institut f\"{u}r Physik, Universit\"{a}t Augsburg, Universit\"{a}tsstra{\ss}e 1, D-86159 Augsburg, Germany}
\affiliation{D.~Ghitsu Institute of Electronic Engineering and Nanotechnologies ASM,Academiei Str. 3/3, MD2028 Kishinev, Moldova}
\affiliation{Present Address: Institute of Applied Physics and Interdisciplinary Nanoscience Center, Universit\"{a}t Hamburg, Jungiusstra{\ss}e 9A, D-20355 Hamburg, Germany}

\author{A.~Ullrich}
\affiliation{Experimentalphysik II, Institut f\"{u}r Physik, Universit\"{a}t Augsburg, Universit\"{a}tsstra{\ss}e 1, D-86159 Augsburg, Germany}

\author{C.~M\"{u}ller}
\affiliation{Experimentalphysik II, Institut f\"{u}r Physik, Universit\"{a}t Augsburg, Universit\"{a}tsstra{\ss}e 1, D-86159 Augsburg, Germany}

\author{A.~S.~Sidorenko}
\affiliation{D.~Ghitsu Institute of Electronic Engineering and Nanotechnologies ASM,Academiei Str. 3/3, MD2028 Kishinev, Moldova}

\author{S.~Horn}
\affiliation{Experimentalphysik II, Institut f\"{u}r Physik, Universit\"{a}t Augsburg, Universit\"{a}tsstra{\ss}e 1, D-86159 Augsburg, Germany}

\author{L.~R.~Tagirov}
\affiliation{Experimentalphysik II, Institut f\"{u}r Physik, Universit\"{a}t Augsburg, Universit\"{a}tsstra{\ss}e 1, D-86159 Augsburg, Germany}
\affiliation{Solid State Physics Department, Kazan Federal University, Kremlevskaya Str. 18, 420008 Kazan, Russian Federation}

\author{A.~Loidl}
\affiliation{Experimentalphysik V, Center for Electronic Correlations and Magnetism, Institut 
f\"{u}r Physik, Universit\"{a}t Augsburg, Universit\"{a}tsstra{\ss}e 1, D-86159 Augsburg, Germany}

\author{H.-A.~Krug~von~Nidda}
\affiliation{Experimentalphysik V, Center for Electronic Correlations and Magnetism, Institut 
f\"{u}r Physik, Universit\"{a}t Augsburg, Universit\"{a}tsstra{\ss}e 1, D-86159 Augsburg, Germany}

\author{R.~Tidecks}
\affiliation{Experimentalphysik II, Institut f\"{u}r Physik, Universit\"{a}t Augsburg, Universit\"{a}tsstra{\ss}e 1, D-86159 Augsburg, Germany}


\date{\today}

\begin{abstract}
We investigated the upper critical magnetic field, $H_{c}$, of a superconductor-ferromagnet (S/F) bilayer of Nb/Cu$_{41}$Ni$_{59}$ and a Nb film (as reference). We obtained the dependence of $H_{c\perp}$ and $H_{c\parallel}$ (perpendicular and parallel to the film plane, respectively) on the temperature, $T$, by measurements of the resistive transitions and the dependence on the inclination angle, $\theta$, of the applied field to the film plane, by non-resonant microwave absorption. Over a wide range, $H_{c\perp}$ and $H_{c\parallel}$ show the temperature dependence predicted by the Ginzburg-Landau theory. At low temperatures and close to the critical temperature deviations are observed. While $H_{c}(\theta)$ of the Nb film follows the Tinkham prediction for thin superconducting films, the Nb/Cu$_{41}$Ni$_{59}$-bilayer data exhibit deviations when $\theta$ approaches zero. We attribute this finding to the additional anisotropy induced by the quasi-one-dimensional Fulde-Ferrell-Larkin-Ovchinnikov (FFLO)-like state and propose a new vortex structure in S/F bilayers, adopting the segmentation approach from high-temperature superconductors.
\end{abstract}

\maketitle 
\section{Introduction}
Singlet superconductivity and ferromagnetism are two antagonistic orders. The formation of singlet Cooper pairs requires electrons with anti-parallel spins, whereas the ferromagnetism tends to align electron spins parallel. Nevertheless, Fulde-Ferrell \cite{FF} and Larkin-Ovchinnikov \cite{LO} (FFLO) predicted superconductivity on a ferromagnetic background, however, in a very narrow range of parameters (see Fig. 22 of Fulde's review\cite{Fulde}). Therefore, only a few experimental realizations exist so far, in heavy fermion and organic superconductors (see the work of Zwicknagl and Wosnitza \cite{Zwicknagel10} for a review).\\
For the heavy fermion system, CeCoIn$_5$, specific heat data\cite{Bianchi03, Radovan03}, thermal conductivity\cite{Capan04}, and penetration depth measurements\cite{Martin05} show evidence for the existence of the FFLO state. In quasi-two-dimensional organic superconductors evidence has been obtained from specific heat data\cite{Lortz07} and magnetic torque studies\cite{Bergk10,Bergk11}. However, a spatial oscillation of the order parameter, which is the main feature of the FFLO state, has not yet been observed directly.\\
In layered organic superconductors, an unusual dependence of the transition temperature on the field direction has been predicted theoretically\cite{Croitoru12,Croitoru12_3,Croitoru12_2,Croitoru12_4,Croitoru13}. It is based on the interplay between the vector potential of a magnetic field (applied parallel to the layered structure), the interlayer coupling, and the nodal structure of the order parameter (and its spatial modulation). These calculations shed new light on the interpretation of the results of experimental investigations\cite{Yonezawa08,Yonezawa08_2} as fingerprints of the FFLO state.\\
In superconductor-ferromagnet (S/F) proximity effect systems, \textit{e.g.} in S/F bilayers, a quasi-one-dimensional FFLO-like state can be realized by Cooper pairs migrating from the superconductor into the ferromagnet \cite{Buzdin05, Eschrig11}. Due to the exchange splitting in the ferromagnet, the Cooper pair gains a non-zero momentum, resulting in an oscillating pairing wave function \cite{Buzdin05, Eschrig11, Tagirov98, LP,BVE}. Its reflection at the outer surface of the F-layer leads to interference effects, yielding a superconducting transition temperature, $T_c$, oscillating as a function of the F-layer thickness, $d_F$ \cite{Tagirov98, Zdravkov06, Sidorenko10}.\\
In the presence of two F-layers (\textit{i.e.} for F/S/F and S/F/F structures), the superconducting transition temperature depends on the relative orientation of their magnetizations \cite{Oh,Tag1}. Such systems represent superconducting spin valves, which can be switched between two states with different transition temperatures by magnetic fields, as demonstrated experimentally for the F/S/F \cite{Gu02,Potenza05,Nowak08} and S/F/F \cite{Nowak13, Leksin10, Leksin11} case.\\
For non-collinear orientations of the magnetizations, an unconventional odd-in-frequency triplet s-wave pairing \cite{BVE} is predicted, reducing the superconducting transition temperature \cite{Fominov10}. Thus, a triplet spin-valve effect \cite{Fominov10} can be established, which could be observed experimentally in S/F/F heterostructures \cite{Zdravkov13,Leksin12} and seems to play a crucial role in a recently realized F/S/F memory element \cite{Zdravkov13_2}. Moreover, in S/F/S Josephson junctions it is possible to realize $\pi$-junctions, in which the phase of the FFLO-like pairing wave function changes by $\pi$ across the device \cite{Ryaz1,Ryazanov01, Oboznov06}. This structure is already applied to fabricate $\pi$-shifters for superconducting digital quantum circuits \cite{Khabipov10,Feofanov10}.\\
Most of these devices are operated by applying a magnetic field to the system. If the S-layer is a type II superconductor (often Nb is used, which is a type II material) vortices appear above the lower critical field. However, also in the case of type I materials, the electron mean free path, $l$, in nanoscale thin film structures may be reduced so far, that the S-layer changes to type II behavior. For example, this is the case for In and Pb at $l=35$~nm and $460$~nm, respectively, calculated using equations and parameters from literature\cite{Tidecks80,Werner86,Werner87}.\\
For Nb it is $\mu_0H_{c1}$= 100~mT and $\mu_0H_{c2}\approx$ 400~mT (at 4.2~K for a polycrystalline rod with $T_{c0}$ = 9.1~K) \cite{Weber91}. Here, $H_{c1}$ and $H_{c2}$ are the lower and upper critical magnetic fields, respectively, $\mu_0$ is the vacuum permeability, and $T_{c0}$ the critical temperature. A detailed study of the temperature dependence of $H_{c1}$ and $H_{c2}$ for Nb is given by Finnemore et al. \cite{Finnemore66}.\\
In S/F structures with Nb as S-material, $H_{c1}$ is very small and, thus, the superconducting layer is soon driven into the Shubnikov phase if a magnetic field is applied. Intuitively, one would expect that the flux quanta penetrating the S-layer also have to be generated (and shielded) in the (superconducting) FFLO-like state in the F-layer.\\
While the vortex state and dynamics in low-$T_c$ and high-$T_c$ superconductors is widely investigated \cite{Blatter94,Huebener01,Huebener02}, there are only a few publications concerning the vortex matter in the FFLO state\cite{Bulaevski03,Ikeda07,Zwicknagel10,Dao13}. In this state a vortex lattice may get pinned at the oscillating FFLO order parameter\cite{Bulaevski03,Uji06,Ikeda07,Zwicknagel10}. While different lattices for the FFLO state have been theoretically proposed \cite{Zwicknagel10,Jiang07,Denisov09}, all of them seem to exhibit nodal planes of the order parameter, as present in the quasi-one-dimensional case, which should be favorable sites for vortex pinning. However, for the quasi-one-dimensional FFLO-like state the vortex state and dynamics is so far unexplored.
\section{Experimental Methods}
The samples of the present work, a thin film S/F bilayer (S=Nb, F=Cu$_{41}$Ni$_{59}$) and a thin Nb film, were deposited by magnetron sputtering on a Si substrate. The thickness of the layers and the composition of the ferromagnetic alloy of the S/F sample, S23\#5, were determined by Rutherford Backscattering Spectroscopy (RBS), yielding $d_S=14.1$~nm and $d_F=34.3$~nm and an alloy composition of $41$~at\% Cu and $59$~at\% Ni. To check the quality and the thickness of the single Nb film, Nb5/1, cross-sectional High Resolution Transmission Electron Microscopy (HRTEM) was applied, resulting in $d_S=7.3$~nm. For details concerning sample preparation and characterization see Appendix A.1.\\
The thin film S/F bilayer of the present work, as well as a Nb reference film, were investigated by measurements of the resistive transitions under applied field and a non-resonant microwave absorption study. To determine the upper critical fields of the samples for fields, applied perpendicular and parallel to the film plane, the superconducting resistive transitions as a function of temperature at fixed magnetic fields were measured in an Oxford Instruments Heliox Sorption Pumped $^3$He Insert (using a lock-in technique with a current of about $50$~$\mu$A at a frequency of $18.792$~Hz). The superconducting transition temperature corresponding to the fixed upper critical field is evaluated as the mid-point of the resistive transition.\\
The angular dependence of the upper critical field at temperatures close to the critical temperature, $T_{c0}$, was investigated by non-resonant microwave absorption. This technique has only been applied so far to study the properties of bulk superconductors \cite{Mamoon5, Mamoon4, Mamoon3, Owens, Shaltiel91, Shaltiel, Shaltiel08, Shaltiel08_2, Shaltiel09, Shaltiel10}. In most cases the Induced Microwave Dissipation by AC Magnetic Field (IMDACMF) technique has been used \cite{Shaltiel91, Shaltiel, Shaltiel08, Shaltiel08_2, Shaltiel09, Shaltiel10}, which we also apply for the measurements of the present work. Sketches of the experimental set up, including the relative orientation of the high frequency (HF) microwave field, the static (DC) magnetic field and the modulating (AC) magnetic field applied, are given by Shaltiel et al. \cite{Shaltiel, Shaltiel08, Shaltiel08_2, Shaltiel09, Shaltiel10}.\\
The basic mechanism of the microwave absorption in this technique is that the magnetic state is defined by the DC magnetic field, whereas the AC modulation tends to reduce the pinning energy of the vortices by 'shaking' them. The 'shaking' occurs, because the AC modulation yields a change of the flux through the sample and, thus, the need of additional or less vortices penetrating the sample and rearrangement of the whole vortex structure. This yields the possibility of vortex motion, resulting in absorption of the high-frequency microwave.\\
Thus, the AC magnetic field induces a modulation signal into the microwave power, $P$, reflected from the cavity\cite{Shaltiel09}. This microwave power is rectified by a diode and fed into a lock-in detector. The signal, $\text{d}P/\text{d}H$ (sometimes also called 'intensity' in literature), detected at the fundamental frequency (also denoted as first AC harmonic in literature) of $H_{AC}$, is obtained from the lock-in detector, \textit{i.e.} information about the microwave dissipation of the sample due to the AC modulation in the state determined by the DC magnetic field is obtained.\\
For details on the IMDACMF technique see Appendix A.2, where also the conversion of magnetic fields from the cgs emu unit system (used in Chap. IV-B) into the international SI system (applied in Chaps. III and IV-A) is given.
\section{Theoretical Framework}
Within the Ginzburg-Landau (GL) theory it can be shown for a thin superconducting film in a parallel magnetic field, $H_\parallel$, that, if the thickness $d$ is smaller than $\sqrt{5}\lambda(T)$ (condition for the transition to the normal state to be of second order), the parallel critical field, $H_{c\parallel}$, is given by\cite{TinkhamBook,Tinkham}
\begin{equation}
H_{c\parallel}(T)=2\sqrt{6}H_{cth}(T)\lambda(T)/d
\end{equation}
In a second order phase transition the superconducting order parameter, $\psi$, of the GL theory\cite{TinkhamBook} (with $|\psi|^2$, representing the density, $n_s$, of the superconducting charge carriers) approaches zero continuously, when $H_\parallel$ is increased to $H_{c\parallel}$. Here, $H_{cth}$ is the thermodynamical critical field of the bulk material\cite{TinkhamBook,Tinkham}. Moreover, $\lambda(T)$ is the penetration depth in weak fields\cite{Tinkham}, which is given by\cite{Tidecks90} $\lambda(T)=0.5^{1/2}\lambda_L(0)\left[T_{c0}/\left(\chi\left(T_{c0}-T\right)\right)\right]^{1/2}$, with \mbox{$\lambda_L^2(0)=3/(2e^2\mu_0N_0v_F^2)$}, where $e$ is the elementary charge, $N_0$ is the number of electronic states (in the free electron model) for one spin direction per volume and energy interval at the Fermi level, and $v_F$ is the Fermi velocity. Furthermore, $\chi=(1+0.752\xi_0/l)^{-1}$, with $l$ the electron mean free path and $\xi_0=\tilde{\gamma}\hbar v_F/(\pi^2kT_{c0})$ the Bardeen-Cooper-Schrieffer (BCS) coherence length, where $\hbar=h/(2\pi)$ with $h$ the Planck constant, $k$ the Boltzmann constant, and $\tilde{\gamma}=\text{exp}(\gamma)=1.781...$ with $\gamma=0.5772...$ the Euler-Mascheroni constant (both $\gamma$ and $\tilde{\gamma}$ sometimes found in literature as Euler constant).\\
Thus, for a thin film with $d\approx l \ll \xi_0$ one obtains just the expression for $\lambda_{eff}(T)$ considered by Tinkham to be appropriate to be used in Eq.(1) [See Tinkham's book\cite{TinkhamBook}, Chap. 4.6, together with Eqs.(3.136) and (3.123), which in the GL regime is given by Eq.(3.123b)].\\
Using the relation\cite{Tidecks90}
\begin{equation}
B_{cth}(T)\xi(T)\lambda(T) 2e = \hbar/\sqrt{2}
\end{equation}
it is possible to rewrite Eq.(1) as
\begin{equation}
H_{c\parallel}(T)=\sqrt{3}\Phi_0/(\pi\xi(T)d\mu_0)
\end{equation}
Here, $\Phi_0=h/(2e)=2.07\cdot10^{-15}$~Tm$^2$ is the elementary flux quantum. Furthermore, $\xi(T)=0.74\chi^{1/2}\xi_0\left[T_{c0}/\left(T_{c0}-T\right)\right]^{1/2}$ is the GL coherence length\cite{Tidecks90} and $B_{cth}=\mu_0 H_{cth}$.\\
For a thin superconducting film in a magnetic field perpendicular to the film plane, Tinkham developed an elementary theory\cite{Tinkham} for the critical field, $H_{c\perp}$. The theory is based on the GL theory and the London theory. It describes the superconducting transition within a model based on the concept of fluxoid quantization. Again, a second order phase transition is assumed. From a detailed discussion of the free enthalpy difference of the superconducting and normal conducting state, the maximum field with a non-vanishing order parameter can be determined, yielding
\begin{equation}
H_{c\perp}(T)=4\pi\mu_0\lambda^2(T)H_{cth}^2(T)/\Phi_0
\end{equation}
Using Eq.(2), this result can be rewritten as
\begin{equation}
H_{c\perp}(T)=\Phi_0/(2\pi\xi^2(T)\mu_0)
\end{equation} 
This is equal to the expression for the upper critical field in bulk samples\cite{TinkhamBook}, \textit{i.e.} $H_{c\perp}(T)=H_{c2}(T)$. Combing Eqs.(5) and (2), we obtain $H_{c\perp}(T)=\sqrt{2}\kappa H_{cth}(T)$, where $\kappa = \lambda(T)/\xi(T)$ is the GL parameter, yielding $\kappa=0.956\lambda_L(0)/(\xi_0\chi)$ using the expressions given above. If the superconducting film is very thin, $l$ is limited by $d$\cite{TinkhamBook} and, thus, $\kappa$ varies with the film thickness.\\
In his elementary theory\cite{Tinkham}, Tinkham also considered the angular dependence of the critical field. From the calculated expression for the free enthalpy density difference, he concludes, that the perpendicular field component leads to an energy term scaling linearly with $H$, while the parallel field component results in a term quadratic in the field, which both have to be balanced against the condensation energy. The origin of this difference is, that the current loops of the perpendicular vortices can scale down, as $H$ increases, while vortices parallel to the thin film are fixed in one dimension by the film thickness.\\
From these arguments Tinkham concluded that for a given angle $\theta$ between the film plane and the magnetic field it is\cite{Tinkham}
\begin{equation}
\left|\frac{H_c(\theta)\text{sin}(\theta)}{H_{c\perp}}\right|+\left(\frac{H_c(\theta)\text{cos}(\theta)}{H_{c\parallel}}\right)^2=1
\end{equation}
Here, for $\theta=0^\circ$ and $\theta=90^\circ$ the field is parallel and perpendicular to the film plane, respectively.\\
In his book\cite{TinkhamBook}, Tinkham pointed out, that the limiting values [given by Eqs.(3) and (5)], as well as his formula for intermediated angles [given by Eq.(6)] are only valid if $d\ll\xi(T)$, so that $|\psi|$ can be regarded as constant over the thickness of the thin film.\\
It is possible to derive the 'Tinkham formula', given by Eq.(6), from the linearized GL equation as a thin film limit (introducing a suitable vector pontential). Details of this derivation are given in Appendix B.1.\\
The GL theory used to derive the results given above is only valid for temperatures just below the critical temperature $T_{c0}$. So far, as the phenomenological GL equations are applied, the results are valid independent of the strength of the electron-phonon interaction. This is the case for Eqs. (1)-(6) and those in Appendix B.1. The explicit expressions for $\lambda(T)$, $\xi(T)$, and $\kappa$, however, resulting from the microscopic derivation of the GL equations by Gorkov, are only valid in the weak coupling limit.\\
The GL theory and the theory of type II superconductors in a magnetic field has been extended to low temperatures. However, no 'Tinkham-like' formula for lower temperatures has been derived. For a detailed discussion, see Appendix B.2.\\
We will apply the theoretical results to our samples, although they are (or contain) films of Nb, which is not a weak coupling superconductor. There is a detailed discussion in Appendix B.3, why this may be allowed.\\
The critical temperature, $T_{c0}$, in the equations above is defined as the superconducting transition temperature, $T_c$, in the absence of currents and magnetic fields. 
Strictly obeying the definition of $T_{c0}$, it can not be defined for S/F heterostructures (therefore it was denoted $T_c$ in our former works), because a magnetic material is present in the sample. Nevertheless, in the present work we identify the transition temperature of S/F heterostructures in zero magnetic field also with $T_{c0}$.\\
\section{Results and Discussion}
\subsection{Temperature Dependence of the Upper Critical Fields}
The temperature dependencies of the upper critical fields perpendicular and parallel to the film plane for the samples S23\#5 and Nb5/1 are shown in Fig. 1 (a) and (b), together with the linear regressions according to Eqs. (3) and (5). The upper critical fields follow the predicted temperature dependencies, \textit{i.e} $\mu_0H_{c\parallel}(T)\propto(1-T/T_{c0})^{1/2}$ and $\mu_0H_{c\perp}(T)\propto 1-T/T_{c0}$ over a wide range of temperatures. Deviations are observed for low temperatures (as expected, because the GL theory should not be valid here) and in the direct vicinity of the critical temperature (possible reasons will be discussed at the end of Chap.~IV-A).\\
Thus, the critical temperatures of the measurement, $T_{c0,MS}=6.42$~K and $6.07$~K for S23\#5 and Nb5/1, respectively, deviate from $T_{c0,GL}$ (simply called $T_{c0}$ in the following) determined by the extrapolation of the temperature behavior of the critical fields predicted by the GL theory.\\
The obtained critical temperatures $T_{c0}($S23\#5$)$ and $T_{c0}($Nb5/1$)$ are $6.34$~K and $5.95$~K, respectively. The slopes of the linear regressions \mbox{$\frac{\text{d}\left(\mu_0H_{c\perp}\right)}{\text{d}T}$} are $-0.316$~T/K and $-0.592$~T/K for S23\#5 and Nb5/1, respectively. For \mbox{$\frac{\text{d}(\mu_0H_{c\parallel})^2}{\text{d}T}$} we obtain $-8.93$~T$^2$/K and $-39.44$~T$^2$/K, respectively. The given values are obtained by a general fit to both the positive and negative field data, obtained for both increasing and decreasing temperature, with a single parameter $T_{c0}$ for both field directions.\\
\begin{figure}
\includegraphics[width=\columnwidth]{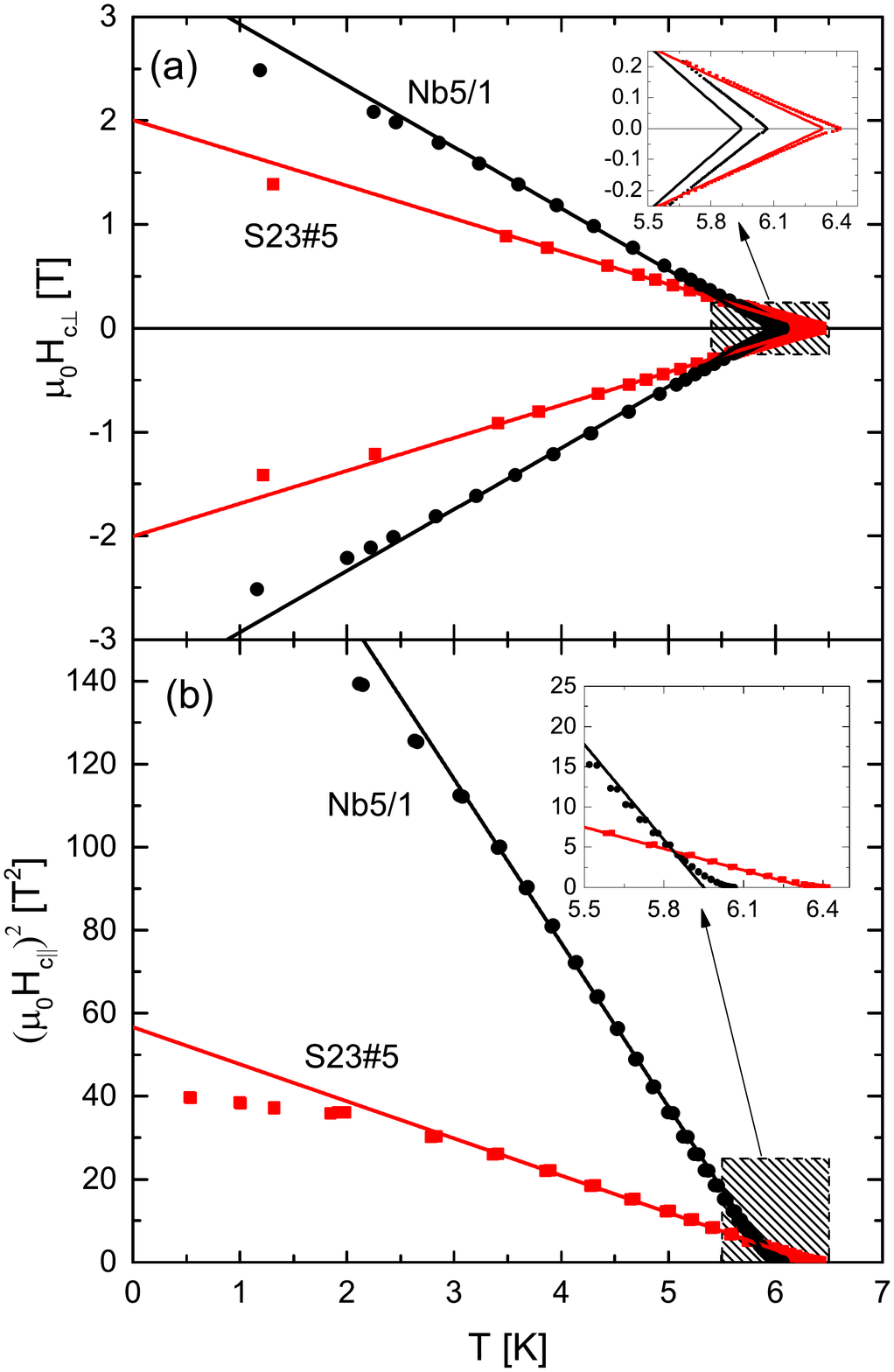}
\caption{(Color online) Temperature dependence of the upper critical fields, $H_{c\perp}(T)$ (a) and $H_{c\parallel}(T)$ (b), perpendicular and parallel to the film surface, respectively, for Nb5/1 (black dots) and the S/F bilayer S23\#5 (red squares). For the parallel field, $(\mu_0H_{c\parallel})^2(T)$ has been plotted.\\The inserts show an enlargement of the data near the critical temperature, $T_{c0}$. The solid lines in (a) and (b) show linear regressions according to Eqs. (5) and (3), respectively.}
\end{figure}
According to Eq.(5), it is
\begin{equation}
\xi(0)=0.74\chi^{1/2}\xi_0=\left[-\frac{2\pi T_{c0}}{\Phi_0}\cdot\frac{\text{d}\left(\mu_0H_{c\perp}\right)}{\text{d}T}\right]^{-1/2}
\end{equation}
Thus, we obtain $\xi(0)= 12.8$ nm and $9.7$ nm for S23\#5 and Nb5/1, respectively. We should emphasize, that while $\xi(0)$ for Nb5/1 is a direct property of the sample, it is in contrast only an effective coherence length, reflecting the whole (inhomogenous) superconducting state in S23\#5.\\
Another part of the same film Nb5 was investigated in our former work\cite{Sidorenko10} by the same measurements presented here. We obtained $T_{c0}=6.25$ K (measured also earlier as $T_{c0}=6.40$ K) and \mbox{$\frac{\text{d}\left(\mu_0H_{c\perp}\right)}{\text{d}T}=-0.558$ T/K}, resulting in $\xi(0)=9.7$ nm. While $T_{c0}$ is somewhat higher, $\xi(0)$ is in good agreement with the value obtained in the present investigation.\\
According to Eq.(3) it is
\begin{equation}
d=\left[-\frac{\pi^2\xi(0)^2T_{c0}}{3\Phi_{0}^2}\cdot\frac{\text{d}(\mu_0H_{c\parallel})^2}{\text{d}T}\right]^{-1/2}
\end{equation}
Inserting the respective quantities for S23\#5 and Nb5/1, we obtain $d=15.6$~nm and $7.7$~nm, respectively. For Nb5/1 the value is very close to $d_S=7.3$~nm, obtained from TEM investigations (see Appendix A.1). For S23\#5 the value is not directly related to the geometry of the sample. These thicknesses represent effective values entering the GL expression for the parallel critical field and, thus, will be referred to as $d_{GL}$ in the following.\\ 
Moreover, in our former work\cite{Sidorenko10}, we also investigated the temperature dependence of the critical fields of a Nb film of $14$~nm thickness (nearly equal to $d_S$ in S23\#5) by the same measurements presented here. We obtained $T_{c0}=8.00$ K (measured also earlier as $T_{c0}=8.05$ K) and \mbox{$\frac{\text{d}\left(\mu_0H_{c\perp}\right)}{\text{d}T}=-0.372$ T/K}, resulting in $\xi(0)=10.5$ nm. If we compare these values with the effective values obtained for S23\#5, we see that the S/F bilayer behaves similar to a thicker layer of a more weakly superconducting material. This is just as expected, because the superconducting layer is weakened by the proximity effect, but the superconductivity can extend into the F-layer.\\
Concerning the deviations of $\mu_0H_{c\perp}(T)$ in Nb5/1 from the GL behavior close to the critical temperature, we refer to Weber et al.\cite{Weber91}, who observed a similar bending up in their $\mu_0H_{c2}(T)$ measurements of Nb bulk samples. They could describe the deviations within the anisotropic Eliashberg theory, considering a mean square anisotropy of electron-phonon interaction and of the Fermi velocity. It is, however, unclear whether the  deviations, observed for $\mu_0H_{c\perp}(T)$ of S23\#5 and $\mu_0H_{c\parallel}(T)$ of S23\#5 and Nb5/1, arise from similar effects.\\
\subsection{Angular Dependence of the Critical Field}
\subsubsection{Experimental Results}
As discussed above, the non-resonant microwave absorption signal $\text{d}P/\text{d}H$ is generated by the motion of the vortices in the superconducting phase. Consequently, the upper critical field $H_{c}$ can be evaluated as the point of vanishing absorption.\\
There are different temperatures mentioned in the captions of Figs. 2-4. First, the setpoint of the temperature controller of the Electron Paramagnetic Resonance (EPR) spectrometer, $T_{SP}$, which is slightly higher than the exact measurement temperature $T_{ME}$. Moreover, to be able to compare $(T_c,H_c)$-points obtained by resistive transitions at constant applied field, $H$, with ones obtained from the EPR measurements at constant temperature, we evaluate a midpoint temperature, $T_{MP}$, to which the EPR spectra correspond (assuming that the same $(T_c,H_c)$-point should be obtained from both methods). The determination of $T_{ME}$ and $T_{MP}$ is described in detail in Appendix A.3.\\
Figure 2 shows selected microwave absorption spectra for S23\#5 at $T_{MP}=4.64$ K, well below $T_{c0}$. The data is recorded from $\theta=0^\circ$ to $\theta=90^\circ$. The transition at $H_{c}$ is well defined and quite smooth and sharp. On the other hand, at this temperature, it is technically not possible to measure the upper critical field for fields applied parallel to the film plane, as it exceeds the limit of the magnet of 16 kOe.\\
However, close to $T_{c0}$ the upper critical field is strongly reduced. Thus, a proper choice of the measuring temperature will reduce the magnetic field below the limit of the magnet. To identify the lowest temperature, at which $H_{c\parallel}<16$ kOe, we evaluated the onset of $\text{d}P/\text{d}H$ with decreasing temperature at a constant field of 15 kOe, applied parallel to the film plane. However, for field sweep measurements at constant temperature on Nb5/1, it was still not possible to establish a measuring temperature corresponding to a $H_{c\parallel}$ below 16 kOe.\\
Moreover, if the measuring temperature approaches $T_{c0}$, the transition is increasingly broadened by the increasing influence of the temperature stability on $H_{c}$ arising from the steep slope of $H_{c\parallel}(T)$ close to $T_{c0}$ due to the square root temperature dependence (see Eq. (3)). The signal is also increasingly noisy, which we attribute to flux flow activation by temperature fluctuations.\\
In the following, the data is always recorded starting from $\theta=0^\circ$ to both, $\theta=-90^\circ$ and, subsequently, to $\theta=+90^\circ$.\\
Figure 3(a) shows a contour plot of the collected data cube of d$P/$d$H$ as a function of $H$ and $\theta$ at $T_{MP}=6.10$~K, close to $T_{c0}$. White color represents a signal below the cut-off value of -115, which represents zero effective signal. A clear cusp at 0$^\circ$ is observed. The white area at low fields around $0^\circ$ is an artifact from phase instability between the AC reference and the measured signal. However, near $H_{c}$ the coupling is reestablished. Unfortunately, it was not possible to obtain an evaluable signal from the sample for some angles, mainly for $\theta>45^\circ$.\\
Figure 3(b) shows the field dependence of $\text{d}P/\text{d}H$ for S23\#5 for selected angles at $T_{MP}=6.10$ K. Basically, $H_{c}$ is given by the mid-point of the observed transition. However, the width of transition (and thus its midpoint) is hard to evaluate as it is veiled by the noise. Nevertheless, the point of vanishing signal can be clearly evaluated (black circles in Fig. 3(b)). This corresponds to the upper end of the superconducting transition at the highest temperature within the temperature stability range.\\
The weak angle-independent signals at $1.8$~kOe and $3.3$~kOe can be assigned to paramagnetic resonances of the cavity background, while the signal at 12 kOe is the paramagnetic resonance of oxygen.\\
\begin{figure}
\includegraphics[width=\columnwidth]{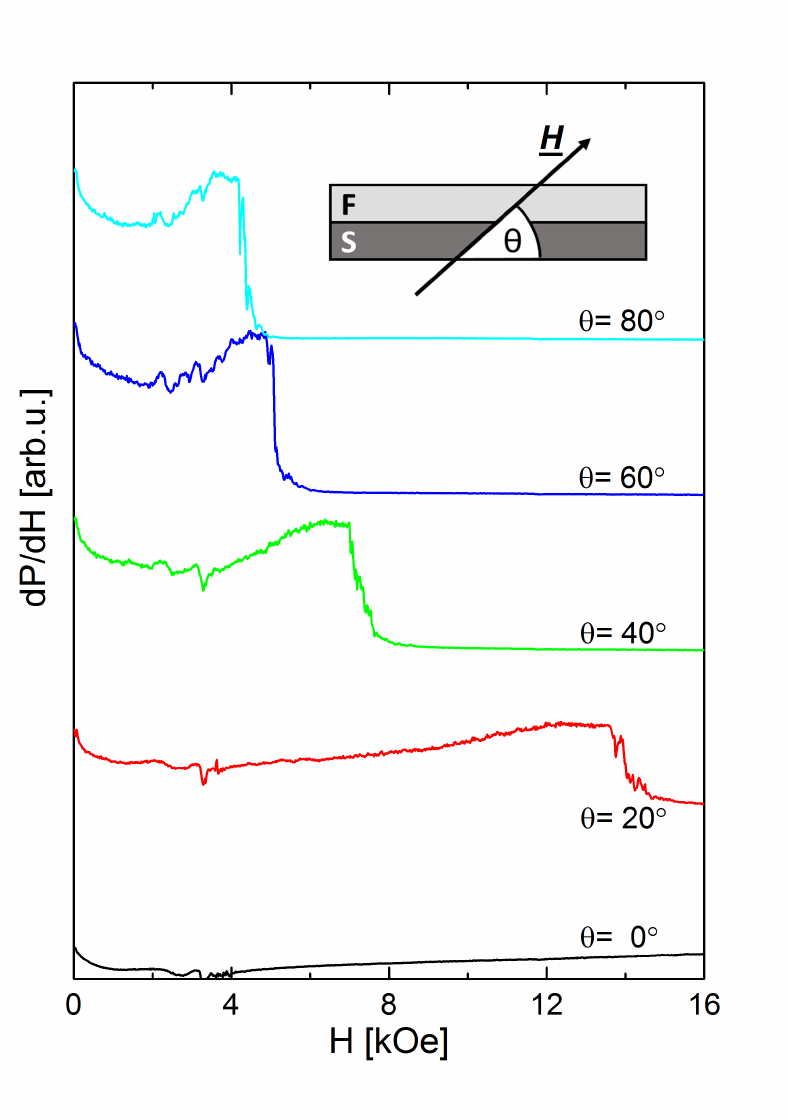}
\caption{(Color online) Selection of microwave absorption spectra at $T_{MP}=4.64$~K for the S/F bilayer, S23\#5, well below the transition temperature, $T_{c0}$, as a function of the applied magnetic field, $H=H_{DC}$, for different angles, $\theta$, between the applied field and the film plane. The individual curves have been offset for better visibility.\\Here, $T_{SP}=5.00$~K and $T_{ME}=4.74$~K (for the definition of the different temperatures see the text, for further details see Appendix A.3). The insert shows a sketch of the sample and definition of the angle $\theta$.}
\end{figure}
\begin{figure}
\includegraphics[width=\columnwidth]{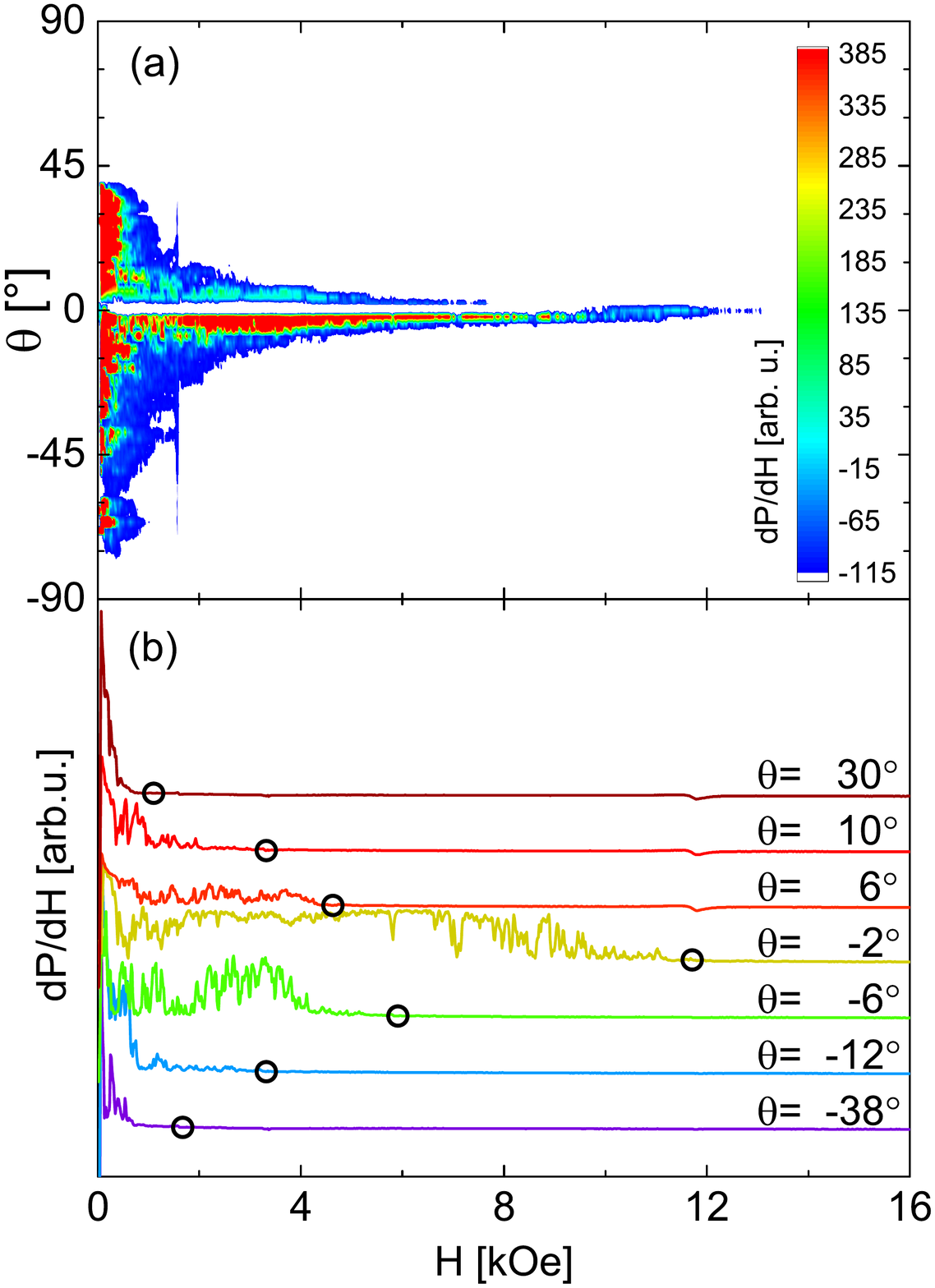}
\caption{(Color online) (a) Contour plot of the raw microwave absorption signal of the S/F bilayer S23\#5 as a function of the applied field, $H=H_{DC}$, and the angle $\theta$, between the applied field and the film plane, at $T_{MP}=6.10$~K, close to the critical temperature, $T_{c0}$. Here, $T_{SP}=6.45$~K and $T_{ME}=6.20$~K. For details see the text and Appendix A.3.\\(b) Microwave absorption spectra, selected from (a) for different angles $\theta$. The individual curves have been offset for better visibility. The black circles show the points of vanishing signal, at which the upper critical field, $H_{c}$, is evaluated. For details see the text.} 
\end{figure}
\subsubsection{Description by the Tinkham Formula}
We applied the Tinkham formula, Eq.(6), to the evaluated angular dependence of $H_{c}$ of both, the reference film Nb5/1 and the S/F bilayer sample S23\#5 close to $T_{c0}$. A detailed analysis of the validity conditions of the Tinkham formula is given in Appendix B.4, showing, that the they are fulfilled for Nb5/1 and, under certain assumptions, also for S23\#5.\\
\mbox{Figure 4} shows the results of the $H_{c}$ evaluation, as well as calculated predictions. The red solid lines in Fig. 4 (a) and (c) are obtained from Eq. (6) by using experimental values from Chap. IV-A (discussed in Appendix A.3). We used $H_{c\perp}(6.10$~K$)=0.882$~kOe and $H_{c\parallel}(6.10$~K$)=15.077$~kOe for S23\#5 and $H_{c\perp}(5.90$~K$)=0.900$~kOe and $H_{c\parallel}(5.90$~K$)=16.300$~kOe for Nb5/1, respectively.\\ 
In contrast to the case of Nb5/1, where the obtained data scatters around the Tinkham prediction, there is a systematic deviation in the case of S23\#5. Thus, the data obtained for S23\#5 has to be discussed in more detail.\\
In Fig. 4 (b) we show predictions obtained from the expressions due to the GL theory for these fields. From Eqs.(3) and (5) (using the results from Chap. IV-A for $T_{c0}$ and $\xi(0)$ and setting $d=d_{GL}$) we obtain $H_{c\parallel}(T)=57.3$ kOe$\cdot (1-T/T_{c0})^{1/2}$ and $H_{c\perp}(T)=20.1$ kOe$\cdot (1-T/T_{c0})$, respectively. Setting $d=d_S$ and $d=d_S+d_F$ yields $H_{c\parallel}(T)=63.4$ kOe$\cdot (1-T/T_{c0})^{1/2}$ and $H_{c\parallel}(T)=18.5$ kOe$\cdot (1-T/T_{c0})^{1/2}$, yielding (using $T=T_{MP}$ and Eq.(6)) the solid and dashed line in Fig. 4 (b), respectively. The case $d=d_{GL}$ is represented by the dotted line.\\
While the general shape of the data is roughly given by the solid line, in particular the data points for angles $|\theta|<40^\circ$ deviate from that prediction. For $|\theta|>10^\circ$ the measurements are better described by the dashed line. Reducing $d$ increases the value of $H_{c\parallel}$, defining the maximum of $H_c(\theta)$, and, thus, the value of $H_c$ around $\theta=0^\circ$. Consequently, the data can be described by changing $d$ (color coded) step by step from $d_S+d_F$ to $d_S$ with decreasing absolute value of the angle. This means, that the value of $H_{c\parallel}$ seems to be determined more and more by the S-layer when $\theta$ approaches zero, \textit{i.e.} the parallel orientation.\\
At a first view, this might indicate that the FFLO-state is weakened or destroyed more and more by the increasing value of the applied field. However, measurements and calculations of the transition temperature oscillations as a function of $d_F$ of F/S/F heterostructures\cite{Antropov13} have shown, that the FFLO-state is neither destroyed nor strongly suppressed for the magnetic fields applied here (at least for high thicknesses $d_{\text{CuNi}}$). To directly compare the results for F/S/F trilayers with those of S/F bilayers, the thickness of the S-layer has to be divided by two\cite{Sidorenko10, Zdravkov11} and only one of the two ferromagnetic layers has to be considered\cite{Zdravkov11} (\textit{i.e.} $d_F$ in the present work has to be compared with approximately $d_{\text{CuNi}}/2$ in our previous work\cite{Antropov13}). See our previous works for details\cite{Sidorenko10,Zdravkov11}.\\
Moreover, also stray fields can be excluded, because above a field of 2-3~kOe the F-layer is in the saturated state \cite{Ruotolo04,Sidorenko10,Kehrle12,Zdravkov13}. There is also no indication of a $T_c$ reduction by stray fields visible in the measurement of $B_{c\perp}(T)$ and $B_{c\parallel}(T)$ presented in Chap.~IV-A. Here, effects of stray fields should be observable at the coercive field of the F-layer (at negative values of $B_{c\perp}\approx-750$~Oe and $B_{c\parallel}\approx -250$~Oe\cite{Kehrle12,Zdravkov13}), where the stray field effects are expected to be most strongly expressed.\\
Thus, this effect seems to result from the special nature of the vortex in S/F bilayers.
\subsection{Vortex State in S/F bilayers}
\subsubsection{Conjecture about a New Vortex Structure}
A possible shape of a vortex in a S/F bilayer, based on the specific anisotropy induced by the quasi-one-dimensional FFLO-like state, optimizing the losses of condensation energy and the energy, needed by the current system to generate the flux quantum and shield the entire superconductor, is proposed in Fig. 5 and will be discussed in detail below.\\
\begin{figure}
\includegraphics[width=\columnwidth]{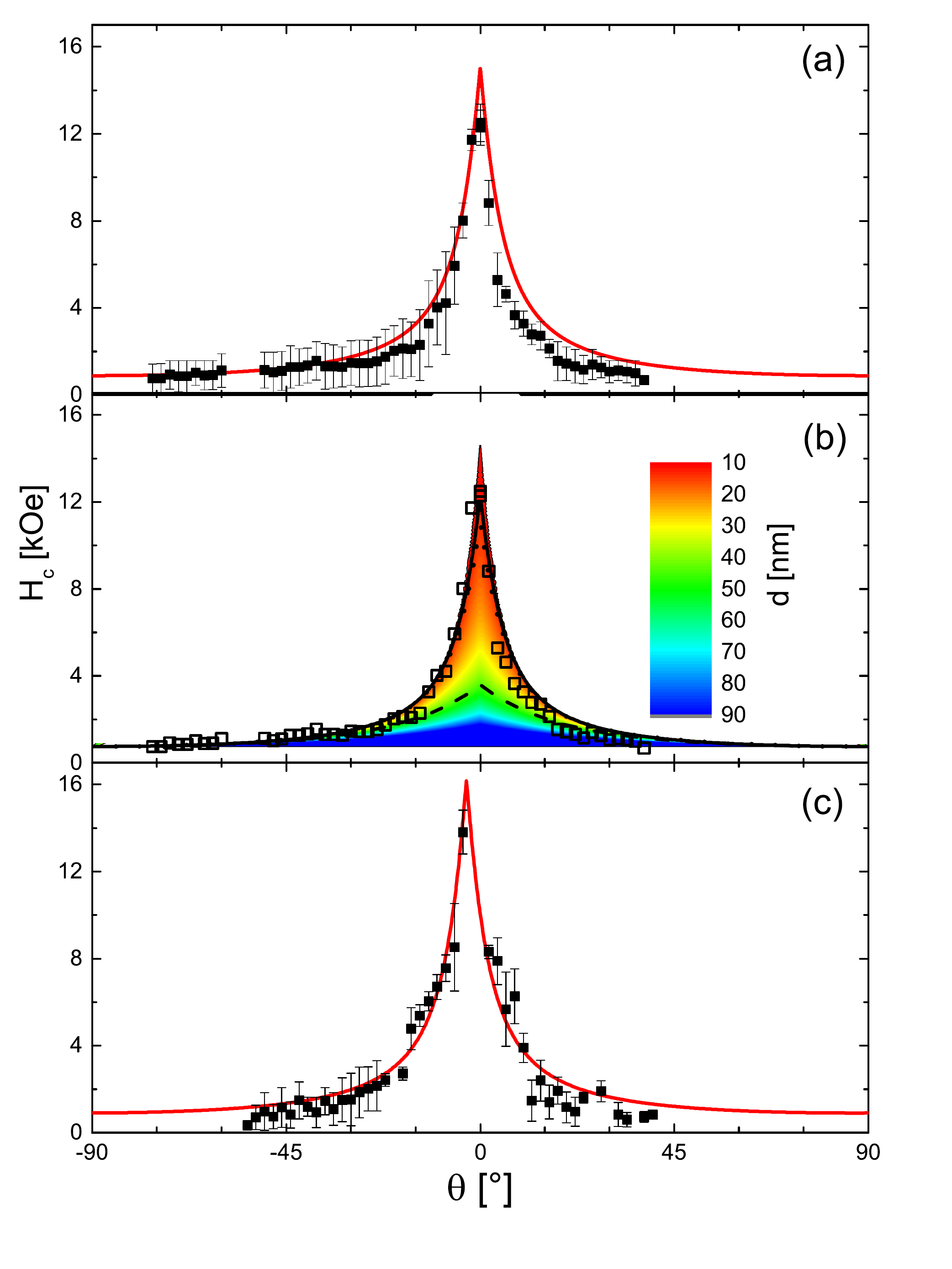}
\caption{(Color online) Upper critical field $H_{c}$ as a function of the angle, $\theta$, between the applied field and the film plane, for the S/F bilayer sample S23\#5 (a) and (b), and the reference film Nb5/1 (c).\\The solid lines in (a) and (c) are descriptions of the data according to Eq.(6), using experimental values for $H_{c\perp}$ and $H_{c\parallel}$, obtained in Chap.~IV-A, at the temperatures $T_{MP}=6.10$~K and $5.90$~K for S23\#5 and Nb5/1, respectively. The values for $T_{ME}$ are $6.20$~K and $6.18$~K, respectively (for details see Appendix A.3). In both measurements $T_{SP}=6.45$~K. The theoretical curve in (c) includes an angular offset, $\Delta\theta=-3.2^\circ$, as the exact angle of the maximum of $H_c$ is not precisely known.\\In (b) the experimental data for S23\#5 is plotted as open squares, while the solid, dashed, and dotted lines represent predictions based on Eq.(6) for different superconducting layer thicknesses, $d=d_S$, $d_S+d_F$, and $d_{GL}$, respectively. Moreover, $H_c(\theta)$ is presented for continuous $d$ (color coded), calculated from Eqs. (3),(5), and (6), to illustrate the decrease of $d$ from $d_{S}+d_{F}$ (dashed line) to $d_S$ (solid line). For details see the text.}
\end{figure}
\begin{figure*}
\includegraphics[width=\textwidth]{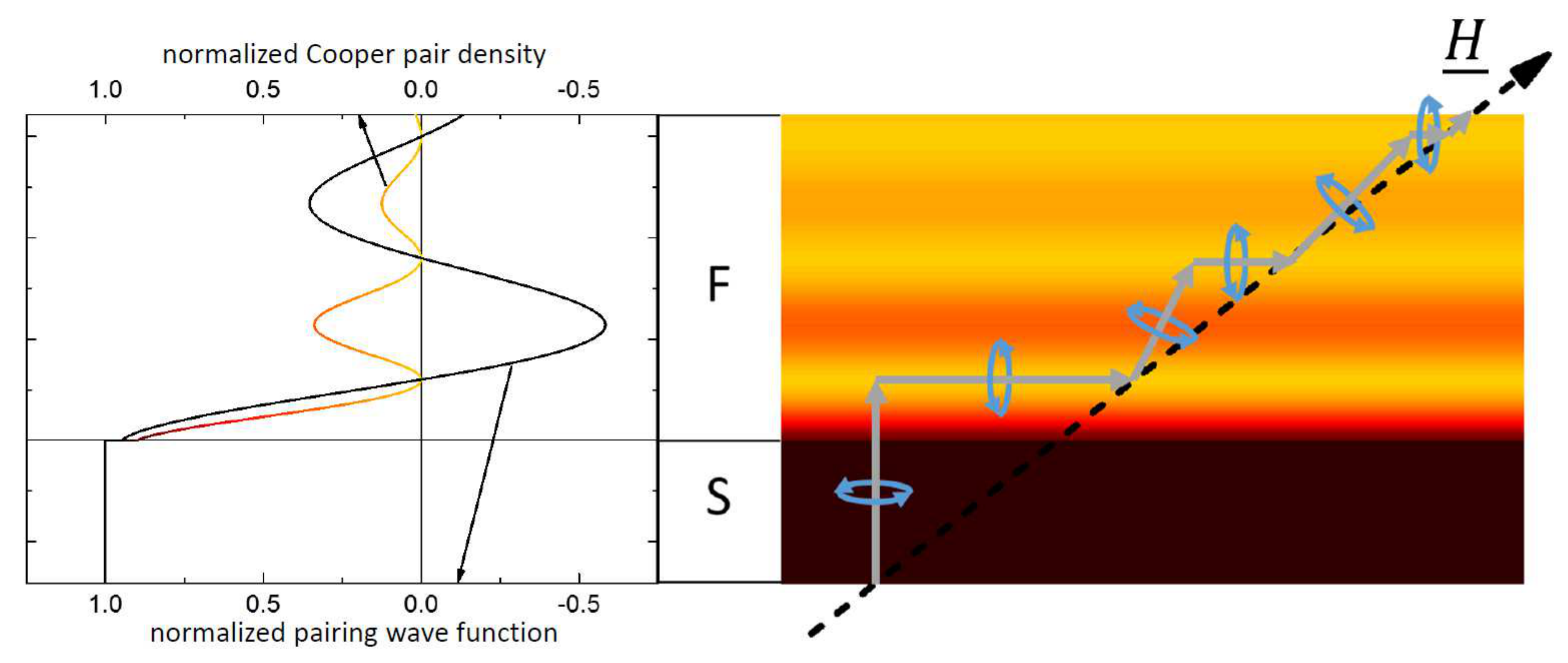}
\caption{(Color online) Left panel: Pairing wave function and Cooper pair density as a function of the space coordinate normal to the film plane in a S/F bilayer.\\Right panel: Sketch of a possible vortex structure taking into account the anisotropic Cooper pair density (color coded) in S/F bilayers.\\For details see the text.}
\end{figure*}
The oscillation of the pairing wave function inside the ferromagnetic layer leads to infinitely thin normal conducting layers parallel to the film, at the position of the nodes of the pairing wave function \cite{Buzdin05,Tagirov98,Sidorenko09} (see Fig. 5 left panel). This may lead to a vortex structure in our S/F bilayers, which has a certain similarity with that one in layered high-$T_c$ superconductors.\\
Within the Lawrence-Doniach theory\cite{TinkhamBook}, Blatter et al. showed in Chap. VIII-A of their review\cite{Blatter94}, that in layered high-$T_c$ superconductors for large angles, $\vartheta$, between the applied field and the ab-plane of such compounds, the vortex is realized by a stack of pancake vortices, which are perpendicular to the ab-plane and generated by current systems in the (strongly superconducting) ab-plane. When approaching smaller angles, \textit{i.e.} if tan$\vartheta<d_{int}/\xi$ (here $d_{int}$  is the spacing between the ab-planes)(see Chap. VIII-A3 of the review of Blatter et al.\cite{Blatter94}), a 'crossover' to a new vortex structure occurs, when the pancake vortices cannot overlap anymore and have to be connected by Josephson vortices, which are parallel to the ab-plane and realized by currents systems perpendicular to the ab-planes.\\
By introducing a vortex into a superconducting material, the superconductor gains the magnetic flux exclusion energy corresponding to one flux quantum, but has to expend the magnetic energy stored in the current system of the vortex, and looses the condensation energy due to the suppression of the order parameter. The energy gain from one vortex is always the same. However, the corresponding energy loss terms in S/F bilayers might (in analogy to the vortex structure in layered high-$T_c$ superconductors) be reduced by a transition from an inclined vortex, parallel to $\underline{H}$, to a series of 'short-link' vortices inclined by an angle between $\theta$ and 90$^\circ$ to the film plane, connected by vortices parallel to the film plane located in the areas of weak superconductivity generated by the FFLO-like state. While the kinetic energy of the shielding currents is increased in such a vortex structure due to the longer flux line, the energy loss due to the destruction of superconductivity inside the vortex is reduced by channeling the vortex through regions of weak superconductivity.\\
It is worthy to mention, that the current systems in Fig. 5 are only sketched schematically as circular. In more detail, we expect that perpendicular to the plane of Fig. 5 (\textit{i.e.} parallel to the film plane) the current system of the vortex segments parallel to the film plane is elongated as in the case of layered superconductors (see Chap. VIII-A1 of Blatter et al.\cite{Blatter94})\\
In Fig. 5, we approximated the pairing wave function to be constant inside the superconductor and to be an exponential decaying cosine function inside the ferromagnet with a step, due to imperfect transparency, at the interface. The local Cooper pair density is proportional to the absolute square of the pairing wave function. The color coding gives the local Cooper pair density from yellow to brown from low to high. Both quantities are plotted as a function of the space coordinate normal to the interface in Fig. 5 left panel.\\
Due to the exponential damping of the oscillation of the pairing wave function with increasing distance from the S/F interface, the reduction of condensation energy loss, which can be obtained by segmenting the vortex, decreases with the distance from the interface as well. For distances large compared to the decay length, the possible reduction of condensation energy loss would be essentially zero, thus, there will be no segmentation. To take into account this decrease in energy gain by segmentation, we propose the inclination angle of the 'short-link' vortices to decrease with increasing distance from the S/F interface.\\
In the case of S/F bilayers the segments parallel to the layers are only similar to Josephson vortices (see Chap. VIII-A1 of Blatter et al. \cite{Blatter94}). Nevertheless, we will calculate the angle, $\theta$, at which the 'crossover' to a segmented vortex mentioned above might occur, for the case that $d_{int}$ is the distance between the S/F interface and the first maximum of the Cooper pair density in the F-material. To calculate $d_{int}$ for S23\#5, we have to calculate the position of the first minimum of the pairing wave function $\Phi_F(x_F)$, \textit{i.e.} the first maximum of $|\Phi_F(x_F)|^2$. Here $x_F$ is the perpendicular distance from the S/F interface. This calculation, together with a general review of the oscillation properties of the pairing wave function in S/F bilayers is given in Appendix~C, yielding $d_{int}=23.9$~nm. Moreover, using $\xi(0)=12.8$~nm and $T_{c0}=6.34$~K, we get for $T=T_{MP}=6.10$~K that $\xi(T_{MP})=65.9$~nm.\\
Thus, we obtain $\text{tan}(\theta)=23.9$~nm~$/$~$65.9$~nm~$=0.363$, yielding $\theta=19.9^\circ$. For smaller $\theta$ a segmentation of the vortex into pancake-like vortices and vortices parallel to the film plane located in the minima of $|\Phi_F(x_F)|^2$ should be possible (within the analogy to the situation in high-$T_c$ superconductors).\\
According to Blatter et al.\cite{Blatter94} for this angular regime new phenomena are expected, whereas for large $\theta$ a continuous description applies. While it is not obvious, at which angles exactly the deviations from the Tinkham formula in our data occur, especially due to the asymmetry of the data, at least the rough angular regime of the deviations seems to fit.\\
Blatter et al. show in Figs. 32 and 33 of their review\cite{Blatter94} the spatial magnetic field distribution for a pancake vortex in a thin superconducting film and a layered superconductor with vanishing Josephson coupling, respectively. The screening currents in the neighboring layers squeeze the magnetic field into the planar direction. Moreover, in Fig. 35 they show a vortex line at small angles $\vartheta$ (corresponding to $\theta$ in the present work), where the core of the Josephson string is fully developed, guiding the magnetic flux between the superconducting layers. In our case, the shielding currents of the vortex guiding the magnetic flux through the minimum of $|\Phi_F(x_F)|^2$ will penetrate into the S-layer.\\
A possible consequence of the proposed vortex structure might be, that for decreasing absolute value of $\theta$, the segments parallel to the S-layer will increasingly dominate. Thus, the shielding properties of the S-layer become more and more important for the value of $H_{c\parallel}$, because the shielding currents will penetrate the S/F interface (yielding $d$ to be more and more governed by $d_S$ - see Fig. 4 (b)). Finally, in the S/F bilayer sample investigated in the present work, for $\theta=0^\circ$, we expect the vortex to have the core in the first node of the FFLO wave function.\\
A detailed justification of the proposed vortex is beyond the scope of the article. To derive a 'Tinkham formula' for S/F bilayers from the GL equation, it would be necessary to include the properties of the quasi-one-dimensional FFLO-like state into the GL equation for the order parameter and find a suitable vector potential, which generates both the applied field and the magnetization. Aside, that it is expected to be difficult to solve this equation, there is the general problem, that a pairing wave function exists in the F-material, but (at least strictly speaking) no superconducting order parameter (similar to the case of the superconductor/normalconductor proximity effect \cite{Deutscher69, Clarke69}).\\
To extend the approach of Tinkham's original work \cite{Tinkham}, when balancing the energy of the shielding current system against the loss of condensation energy, one has to consider both, the anisotropy of the magnetic and the superconducting state, in the free enthalpy terms, yielding much more complicated and space dependent equations. Moreover, the straight-forward superposition of the parallel and perpendicular enthalpy terms to obtain the Tinkham formula will most probably not be possible.\\
\subsubsection{Vortex Dynamics}
The vortex structure proposed has a pinning behavior, which is expected to be different from that one of a continuous vortex considered, to derive the angular dependence of $H_c$ by Tinkham\cite{Tinkham}. It is expected to behave more similar to the vortices in high-$T_c$ superconductors. Here, the Josephson vortices are much more strongly pinned than the pancake vortices. Thus, the ESR signal should decrease with the increase of the parallel field component. However, there is a mechanism leading to increased Josephson vortex mobility at very small angles\cite{Weidinger97}. While we see the general decrease of the signal amplitude with decreasing angle, a possible stabilization around $\theta=0^\circ$ can not be stated based on the obtained data. Moreover, since the direction of flux movement is not determined by an applied current in contrast to the work of Weidinger et al.\cite{Weidinger97}, the parallel flux through the sample can be changed by moving the vortices only along the nodal plane of the pairing wave function, so it is questionable if such a mechanism is applicable in our case.\\
Investigations of high-$T_c$ superconductors, using the IMDACMF method, with $H_{DC}$ perpendicular to the ab-plane of the layered structure show a non-vanishing signal $($d$P/$d$H)(T)$ at constant field only just below $T_c$, where thermal activated flux flow (TAFF) governs the motion of pancake vortices. For $H_{DC}$ parallel to the ab-planes, where only Josephson vortices are present, the signal intensity increase sharply at $T_c$ with a further increase down to low temperatures (see Fig. 4 of Shaltiel et al.\cite{Shaltiel09}). This leads to the conclusion, that in those materials at temperatures well below $T_c$ the induced microwave dissipation results from the interaction of the microwaves with the Josephson vortices\cite{Shaltiel09}.\\
This is different in the experiments in the present work. For the S/F bilayer we investigated $($d$P/$d$H)(T)$ at constant field for $H_{DC}$ parallel and perpendicular to the film plane. In both cases the signal has its largest value just below $T_c$ and decays to lower temperatures, but does not vanish (at least down to $4$~K).\\
\subsection{Comparison with Related Systems}
\subsubsection{Bulk FFLO Superconductors}
The angular dependence of the upper critical field for the FFLO state in bulk superconductors has been investigated by Dao et al.\cite{Dao13}, considering the role of crystal anisotropy on the vortex state. Contrary to conventional superconductivity, where only the crystal structure influences the type of the Abrikosov vortex lattice, the modulation of the order parameter in the FFLO phase has an influence on the vortex structure, too. In special situations, higher Landau level (LL) states lead to an angular dependence of $H_c$ with transitions between the higher LL states. If only one state is considered, a smooth $H_c(\theta)$ dependence is predicted (between $90^\circ$ and $30^\circ$). In the general case, transitions between different LL states lead to structures in the $H_c(\theta)$ dependence, for $|\theta|<20^\circ$. These structures, however, only appear for low temperatures. Above $T=0.5T_{c0}$ the structures vanish. The overall shape of the $H_c(\theta)$ curve is rounded at $\theta=0^\circ$, \textit{i.e.} it has a shape similar to the Lawrence-Doniach behavior mentioned above. The experiments of the present work are performed closer to $T_{c0}$, where no LL state transition generated structures are predicted. Moreover, we observe a sharp cusp-like behavior of $H_c(\theta)$ in our experiments.\\
\subsubsection{Thin Films and Layered Superconductors}
The angular dependence of the critical field was widely investigated for thin films\cite{Tinkham64,Harper68,Sidorenko81,Banerjee84}, multilayers \cite{Banerjee83,Banerjee84,Chun84,Sidorenko91,Sidorenko96,Klemm2012}, including fractal geometries\cite{Sidorenko96}, and high-$T_c$ superconductors\cite{Palstra88,Naughton88,Juang88,Iye90,Marcon92}. Although the experiments often follow the general behavior of Tinkham's prediction, deviations are found in detail, as observed in the present work. In multilayers and high-$T_c$ superconductors also a Lawrence-Doniach behavior of $H_c(\theta)$\cite{TinkhamBook}, which describes an anisotropic three-dimensional multilayer superconductor, is observed. We ascribe the observed deviations in the present work to a special vortex structure, generated by the quasi-one-dimensional FFLO-like state in the F-material of the S/F bilayer.\\
Since this vortex structure is related to that one of a layered superconductor (see Chap. VIII-A of the review of Blatter et al.\cite{Blatter94}), this conclusion is supported by the investigations of Prischepa et al. \cite{Prischepa05}, who found an angular dimensional cross-over of $H_c(\theta)$ at fixed temperature for Nb/Pd multilayers. Samples of odd and even numbers of normal/superconducting (N/S) bilayers of Pd/Nb (9 and 10, respectively, plus a capping Pd layer) were measured  in the temperature range $T<T^*<T_{c0}$, where the square-root behavior of $H_{c\parallel}(T)$ indicates a two-dimensional behavior (for $T^*<T<T_{c0}$ a linear temperature law of $H_{c\parallel}(T)$ is observed, indicating three-dimensionality). Strong deviations from the Tinkham formula are obtained for $H_c(\theta)$ of the multilayer with an even number of bilayers. Only certain ranges of $H_c(\theta)$ follow Tinkham's prediction. For small angles, however, with $H_{c\perp}$ as a free parameter and for larger angles, with $H_{c\parallel}$ as the free parameter. The latter range is interpreted as an unusual three-dimensional mode, because the large angle tail can be described by the Lawrence-Doniach model.\\
The physical interpretation proposed is that for small angles the superconducting nucleus is localized in one period of the S/N structure, but for large angles, it is spread over more than one period by the perpendicular component of the external magnetic field, resulting in an object with a three dimensional feature (for a detailed discussion see the work of Prischepa et al.\cite{Prischepa05}). It is argued that, nevertheless, the Lawrence-Doniach description is not applicable, because it was deduced in the approximation of the homogeneous infinite medium. The applicability of the Tinkham formula is concluded to be the consequence of a relatively homogeneous order parameter in one S-layer.\\
This is very similar to our S/F bilayer, where the pairing wave function oscillates in the F-material due to the FFLO state, but is nearly constant in the S-layer. Allthough a S/F bilayer is not a multilayer, the oscillating pairing wave function generates strongly and weakly superconducting regions in the F-material. The perpendicular component of the external field shifts more and more 'weight' of the vortex into these strongly superconducting regions for increasing angle. Since our S/F bilayer is always in the two-dimensional regime, a Lawrence-Doniach model can not be applied to explain the measurements, but the Tinkham formula with a changing $H_{c\parallel}$, possibly arising from a varying effective superconducting thickness $d$ increasing from $d_S$ to $d_S+d_F$ for increasing angle, gives a reasonable description in the angular regime, where the segmentation of the vortex is expected.  
\section{Conclusions}
In summary, in a S/F bilayer and a thin Nb film, we investigated the temperature dependence of $H_{c\perp}$ and $H_{c\parallel}$ by measurement of resistive transitions, and the angular dependence of $H_c$ by a non-resonant microwave absorption study.\\
Over a wide temperature range, the temperature dependence of $H_{c\perp}$ and $H_{c\parallel}$ follow the respective linear and square-root behavior, predicted by the Ginzburg-Landau theory. However, close to the critical temperature deviations are observed, and compared to those arising from anisotropies of the electron-phonon interaction and Fermi velocity in niobium.\\
We analyzed the results of the angular dependence of $H_c$ within the framework of Tinkham's theory of thin superconducting films. While the thin Nb film could be well described by this theory, the S/F bilayer data shows deviations at low inclination of the applied field to the film plane.\\
Based on the oscillations of the pairing wave function inside the ferromagnetic layer, induced by the quasi-one-dimensional FFLO-like state, and adopting the approach of a segmented vortex, as present in layered high-$T_c$ superconductors, we propose a new vortex structure, which reduces the energy in the system by alternating steps of vortex short-links through strongly superconducting regions and flux channeling through the weak superconducting minima of the Cooper pair density. Since the pairing wave function is damped as a function of the distance to the S/F interface, the Cooper pair density difference between strong and weak superconducting regions, and thus the possible energy gain, decreases. Therefore, we propose the inclination angle of the short-link part of the vortex to decay with increasing distance from the S/F interface.\\
Although the vortex structure proposed has some similarity to the segmented vortex in high-$T_c$ superconductors, the vortex dynamics seem to be different.\\
Moreover, we discuss our findings and interpretations in the context of investigations of an angular dimensional cross-over of the upper critical field in superconductor-normalconductor multilayers, where special vortex states are discussed to arise from the layered geometry.\\
While there are theoretical studies of the angular dependence of the upper critical field for the FFLO state in bulk superconductors, considering the influence of the modulation of the order parameter on the vortex state, there are so far no such calculations for the quasi-onedimensional FFLO-like state in S/F proximity effect systems. However, such theoretical considerations are strongly desirable, because an extension of the Tinkham formula to this situation seems not to be possible.
\section*{Acknowledgments}
The authors are grateful to S.~Heidemeyer, B.~Knoblich, and W.~Reiber for TEM sample preparation, to J.-M.~Kehrle for taking the TEM image, to W.~Reiber and S.~Gsell for the RBS measurements, and to G.~Obermeier and R.~Horny for technical support concerning the low temperature resistive measurements.\\
This work was supported by the Deutsche Forschungsgemeinschaft (DFG) under the Grant No. HO 955/9-1. R.M. and L.R.T. were supported in part by the Program of Competitive Growth of Kazan Federal University. V.I.Z. was partially supported by ERC advanced grant 'ASTONISH'. The IMDACMF investigations (A.L. and H.-A. K.v.N.) were partially supported by the Deutsche Forschungsgemeinschaft (DFG) within the Transregional Collaborative Research Center TRR 80 'From Electronics Correlations to Functionality' (Augsburg, Munich).\\
\section*{Appendix}
\subsection{Experimental Techniques}
\subsubsection{Sample Preparation and Characterization}
The S/F bilayer sample, investigated in the present work is part of a Nb/Cu$_{41}$Ni$_{59}$ thin-film sample series (S23) produced by magnetron sputtering at room temperature\cite{Sidorenko10}. All targets used in the preparation were first pre-sputtered for 10-15 minutes to remove possible contaminations. Afterwards, an $1$~mm thick commercial \{111\} silicon substrate (size $7$~mm x $80$~mm) was covered with an amorphous silicon buffer layer by RF sputtering to provide a clean surface for the subsequent layers. In the next step a thin niobium layer was produced by applying the 'spray technique' \cite{Zdravkov06,Sidorenko10,Sidorenko09}. In this technique the Nb target was continuously moved across the substrate during the DC sputtering process to ensure a layer of constant thickness and precise control of the growth rate, resulting in a flat niobium layer of thickness $d_S=14.1$~nm.\\
Subsequently, a wedge-shaped ferromagnetic layer was RF sputtered from a Cu$_{40}$Ni$_{60}$-alloy target by using the intrinsic spatial gradient of the deposition rate inside the chamber. Finally, to prevent degradation of the samples under atmospheric conditions, an amorphous silicon cap layer of about 5-10 nm thickness was deposited on top of the sample.\\
Individual samples were then cut perpendicular to the wedge gradient (36 slices, enumerated from the thick to the thin end of the wedge, usual width about $2.5$~mm) from the obtained layered structure. Due to the small thickness gradient of the wedge-like ferromagnetic layer and the small sample width, the thickness of the S and F-layer is regarded constant within each individual sample.\\
We have chosen the sample S23\texttt{\#}5 (size $4.4$~mm x $2.8$~mm), with $d_F=34.3$~nm. In this range of thicknesses, $T_{c0}$ becomes almost independent of $d_F$ - it is $T_{c0}~\approx$~6.4~K, for $d_F >23$~nm and, thus, for sample S23\texttt{\#}5. This does \emph{not} mean that interference effects of the pairing wave function in this range of thicknesses are absent, but only that the interference modulation of the flux of the pairing wave function through the S/F interface is too weak to influence the superconducting state in the whole S-layer. This statement can be justified by the behavior of $T_{c0}(d_F)$ for lower $d_S$ (\textit{e.g.} sample series S21 in our former work\cite{Sidorenko10}) where at comparable thicknesses $d_F$ interference effects are still observable. For details of the argumentation above, see Fig. 6 and Chap. IV of our former work\cite{Sidorenko10}. However, the interference effects are expected to decay as the amplitude of oscillation of the pairing wave function at the outer F-boundary decays with increasing $d_F$. Thus, the pairing wave function in the F-layer is more and more close to the one induced by the underlying FFLO proximity effect. Moreover, we expect the anisotropy of the pairing wave function inside the F-layer to be larger for constructive than for destructive interference (see Fig. 2 of our article on interference effects in S/F bilayers\cite{Sidorenko09}). Thus, we have chosen a relatively thick sample of the series, which should exhibit constructive interference.\\
To distinguish the effects, arising from the influence of the ferromagnetic layer, from those intrinsic to a thin niobium layer, a reference film (Nb5) was produced by the same deposition procedure, however, without the ferromagnetic layer on top, \textit{i.e.} a single niobium layer with constant thickness, which is sandwiched between the amorphous silicon buffer and cap layers. We cutted several parts from Nb5 for different measurements. The part Nb5/1 (size 7 mm x 4 mm) is used for low temperature measurements.\\
\begin{figure}
\includegraphics[width=\columnwidth]{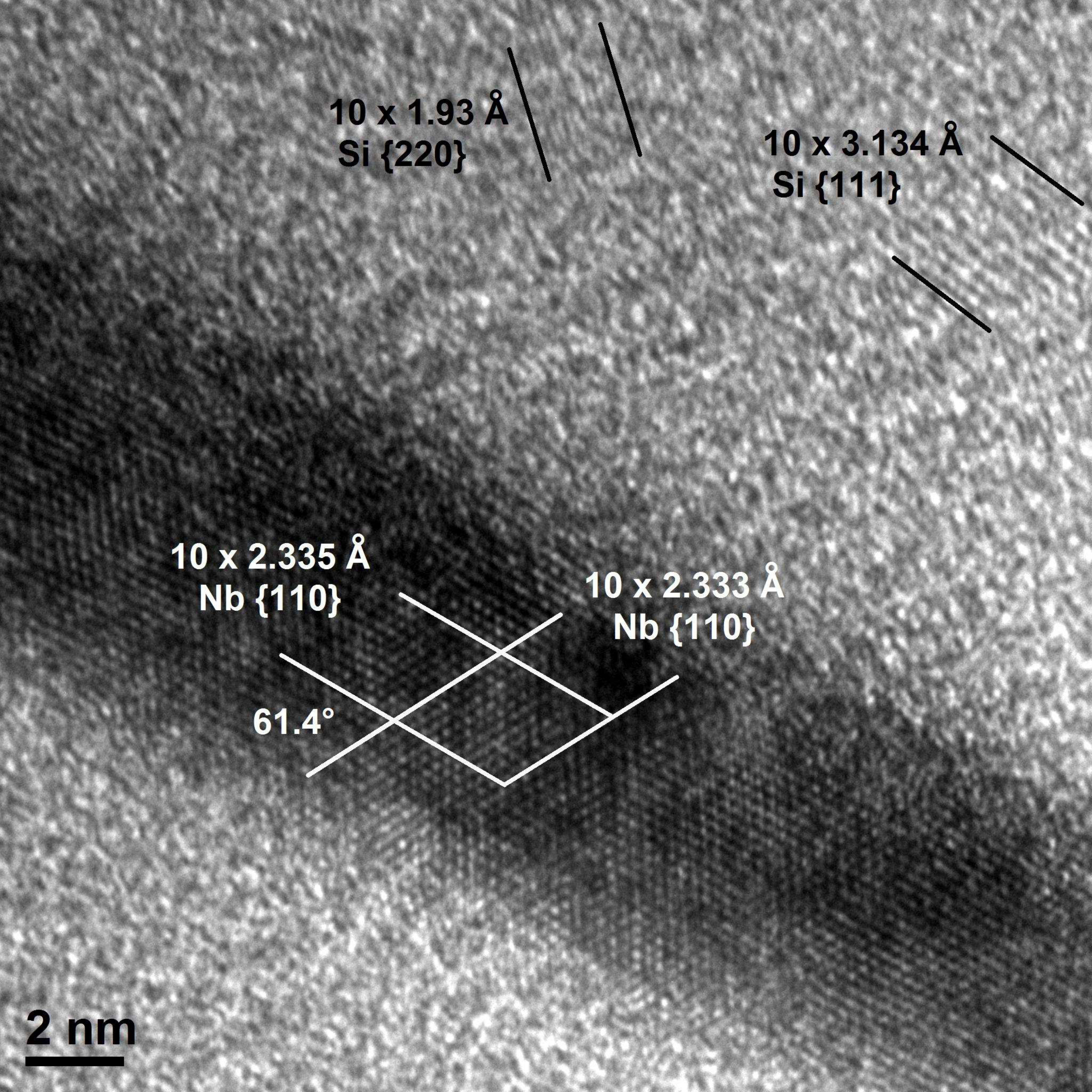}
\caption{Cross-sectional HRTEM image of a part of the niobium film Nb5. The dark niobium layer shows a highly crystalline structure with Nb \{110\} planes. On the upper right side, lattice planes of the silicon substrate are visible and were used to confirm the scale.}
\end{figure}
One part of the reference film, Nb5, was subjected to cross-sectional High-Resolution Transmission Electron Microscopy (HRTEM) to check the thickness and quality of the layer. The cross-section specimen was prepared by conventional dimpling and ion thinning. The obtained HRTEM image is shown in Fig. 6. On the upper right side, the \{111\} planes of the silicon substrate can be clearly identified by their lattice constant of 3.134 \AA. A careful inspection also reveals the Si \{220\} planes with a spacing of 1.93 \AA, including an angle of $\approx$ 35$^\circ$ with the \{111\} planes which is in agreement with the theoretical value of 35.26$^\circ$.\\
The niobium layer is clearly visible due to the strong Z contrast to the silicon substrate. It shows a highly crystalline structure with lattice plane distances of 2.33 \AA. These distances can be attributed to the Nb \{110\} planes. The angle between these planes is $\approx$ 61.4$^\circ$, which corresponds to the theoretical value of 60$^\circ$. Therefore, the viewing direction in the niobium layer could be identified as [111].\\
Lattice types and constants are taken from literature\cite{Handbook}, angles and spacings between lattice planes are calculated according to the known crystal structure.\\
The thickness of the niobium layer is evaluated to be $d_S=$ 7.3 nm. We regard this value to be more accurate than the value of 6.8 nm, obtained by Rutherford Backscattering Spectroscopy (RBS) on Nb5/1 in \mbox{Chap. III-A} of our former work \cite{Sidorenko10}.\\
\begin{figure}
\includegraphics[width=\columnwidth]{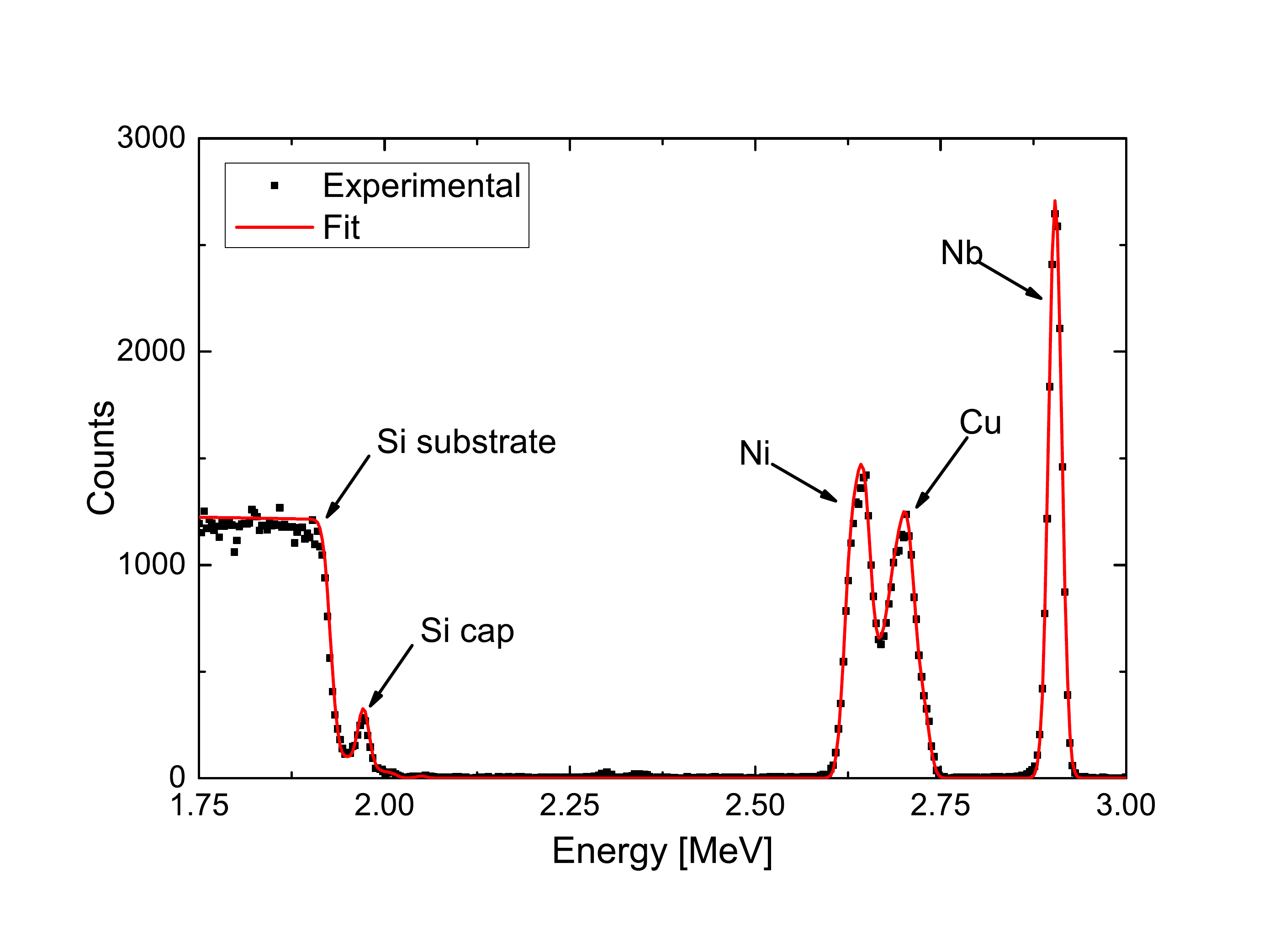}
\caption{(Color online) RBS spectrum of the S/F bilayer S23\texttt{\#}4 (thicker next to S23\texttt{\#}5).}
\end{figure}
To determine the thicknesses and composition of the layers in the S/F bilayer sample we performed RBS with $\alpha$-particles at an energy of 3.5 MeV. However, we did not use S23\texttt{\#}5 to prevent altering of its properties by radiation damage. Instead, we investigated a subset of samples across the whole sample series and obtained the data of S23\texttt{\#}5 by linear interpolation of the results of S23\texttt{\#}4 and S23\texttt{\#}7. Figure 7 shows the RBS spectrum of S23\texttt{\#}4 together with the fit. The fit is in good agreement with the experimental data. The small unfitted feature at approximately 2.35 MeV is an artifact, arising from the sample holder. The different peaks are assigned to the corresponding layers. The shown spectrum is representative for all obtained spectra. For S23\texttt{\#}5 we obtain 14.1 nm and 34.3 nm for $d_S$ and $d_F$, respectively, and a composition of $41$~at\% copper and $59$~at\% nickel for the F-layer.
\subsubsection{Induced Microwave Dissipation by AC Magnetic Field (IMDACMF) Technique}
The non-resonant microwave absorption experiments, presented in Chap.~IV-B, have been performed in a Bruker ELEXYS 500 X-band EPR spectrometer. The microwave source feeds a 9.3~GHz rectangular H102 (also known as TE102) cavity. The sample is positioned at its center, where only the magnetic component of the microwave field is present. Moreover, the sample is exposed to collinear DC ($H_{DC}$) and AC ($H_{AC}$, amplitude 30~Oe, frequency~100 kHz) magnetic fields, applied perpendicular to the magnetic microwave field. The sample can be cooled to low temperatures using a continuous helium flow cryostat (ESR900, Oxford Instruments). The relative orientation of the magnetic DC and AC field with respect to the film plane of the sample (thin films on Si substrate) can be varied using a goniometer. The rotation axis is parallel to the microwave magnetic field (to keep the microwave magnetic field strength unchanged, see sketch in Fig. 1 in the work of Shaltiel \cite{Shaltiel09}) and the long side of the sample. This means for the S/F bilayer investigated in the present work, that it is rotated around the magnetically semi-easy axis\cite{Kehrle07} of the F-layer from its hard axis to its easy axis, which are parallel and perpendicular to the film surface, respectively\cite{Ruotolo04,Kehrle07,Zdravkov13}.\\
The samples were first zero-field cooled from room to liquid helium temperature. Initially, by comparing data obtained in both sweep directions of the magnetic field, we verified that the field sweep direction has nearly no influence on the signal. The data presented in \mbox{Chap.~IV-B} was recorded by sweeping the magnetic field from 0 to 16 kOe at a given angle. After reducing the field to zero, the angle was changed and the procedure was repeated again.\\
Since the Bruker EPR spectrometer used is calibrated in the cgs emu unit system, the applied magnetic fields are measured in Oe. In contrast, the theory in the present work and the results in Chap.~IV-A are given in the international SI system. To convert magnetic fields into the SI system the relation \mbox{$1$~Oe~$=10^3/(4\pi)$~A/m~$=79.58$~A/m} \cite{Goldfarb85} is used. Furthermore, with \mbox{$\mu_0=4\pi \cdot 10^{-7}\frac{\text{Vs}}{\text{Am}}$} one obtains the magnetic flux density related to $1$~Oe as $B=\mu_0H=10^{-4}$~Vs/m$^2$, \textit{i.e.} \mbox{$10$~kOe} yield \mbox{1 T}.
\subsubsection{Calibration of the IMDACMF Temperature Scale}
Usually, the sample temperature measured in the EPR spectrometer is found to be somewhat lower than the setpoint of the temperature controller, $T_{SP}$. Thus, the exact measurement temperature, $T_{ME}$, has to be calibrated. For this purpose, we compare the upper critical fields, $H_{c\perp}$ and $H_{c\parallel}$, obtained by IMDACMF (see Fig. 4), with the data for $H_{c\perp}(T)$ and $H_{c\parallel}(T)$, obtained in \mbox{Chap.~IV-A}, and their respective linear interpolations. The values, which lead to the best description (according to Eq.(6)) of the $H_c(\theta)$ data in Fig. 4 and the related temperatures are \mbox{$H_{c\perp}(5.90$~K~$)=0.900$~kOe} and \mbox{$H_{c\parallel}(5.90$~K~$)=16.300$~kOe} for Nb5/1 and \mbox{$H_{c\perp}(6.10$~K~$)=0.882$~kOe} and \mbox{$H_{c\parallel}(6.10$~K~$)=15.077$~kOe} for sample S23\#5. To obtain these values for Nb5/1, we allowed the parallel alignment to deviate from $\theta=0^\circ$ by $\Delta\theta=-3.2^\circ$, as the exact angle at which the maximum of $H_c$ occurs is not precisely known (for S23\#5 it is $\Delta\theta=0^\circ$).\\
However, the obtained temperatures are \textit{not} the actual measurement temperatures, $T_{ME}$, but the mid-point temperature, $T_{MP}$ of resistive transitions, which would lead to the same data points, if the results would have been obtained from resistive transitions at constant fields. As noted before, the data points are determined by evaluating the vanishing of the IMDACMF signal, when the sample enters the normal state. Consequently, $T_{ME}$ is higher than $T_{MP}$ by half of the transition width at constant field (about \mbox{$500$~mK~$/2 = 250$~mK for Nb5/1} and \mbox{$150$~mK~$/2 = 75$~mK for S23\#5)} and half (about \mbox{50 mK /2 = 25 mK)} of the temperature stability range of the EPR spectrometer. Thus, $T_{ME}=6.20$~K and $6.18$~K for the measurements of S23\#5 and Nb5/1  in Fig. 4 (a) and (c), respectively. In both cases, the setpoint temperature was $T_{SP}=6.45$~K, yielding an average difference of $0.26$~K between $T_{SP}$ and $T_{ME}$. Assuming the same temperature offset also for the measurements on S23\#5 at lower temperatures (see Fig.~2), for which $T_{SP}=5.00$~K, we estimate a corresponding $T_{ME}$ of $4.74$~K and (considering the transition width and temperature stability mentioned above) $T_{MP}$ to be $4.64$~K.\\
 
\subsection{Details on the Theoretical Framework}
\subsubsection{Derivation of the Tinkham Formula from the Linearized Ginzburg-Landau Equation}
The 'Tinkham formula' given in Eq.(6) can be derived from the linearized GL equation, obtained by neglecting the term proportional to $|\psi|^2\psi$ in the GL equation for the order parameter\cite{Tidecks90}, yielding
\begin{equation}
(1/2m')(-i\hbar\underline{\nabla}-e'\underline{A})^2\psi+\alpha\psi=0 
\end{equation}
Here, $m'=2m$ is twice the electron mass, $e'=-2e$ twice the electron charge, $\underline{A}(\underline{r})$ the vector potential of the magnetic flux density with $\underline{B}(\underline{r})=\text{rot}(\underline{A}(\underline{r}))$, and $\alpha=-B_{cth}^2/(\mu_0n_{s0})$ with $n_{s0}=|\psi_0|^2$ the density of the particles described by $\psi$ in the absence of currents and magnetic fields.
Introducing the GL coherence length $\xi^2=-\hbar^2/(2m'\alpha)$ and $\Phi_0=h/(2e)$ one obtains
\begin{equation}
\left[\left(\frac{\underline{\nabla}}{i}-\frac{2\pi\underline{A}}{\Phi_0}\right)^2-\frac{1}{\xi^2}\right]\psi=0
\end{equation}
By choosing a coordinate system, in which $x$ is measured normal to the film from its midplane and a magnetic field lying in the xz-plane, the magnetic field is given by \mbox{$\underline{H}=H(\hat{\underline{x}}\text{sin}(\theta)+\hat{\underline{z}}\text{cos}(\theta))$} with $H=|\underline{H}|$. For a second order phase transition, in a first approximation, the magnetization $\underline{M}$ of a superconductor can be neglected in the direct vicinity of the critical magnetic field, so that $\underline{B}=\mu_0(\underline{H}+\underline{M})\approx\mu_0\underline{H}$. Thus, a vector potential with only a y-component corresponding to this field can be chosen as
\begin{equation}
\underline{A}(\underline{r})=\mu_0H(x\text{cos}(\theta)-z\text{sin}(\theta))\hat{\underline{y}}
\end{equation}
Inserting $\underline{A}$ from Eq.(11) into Eq.(10) yields a differential equation, which is hard to solve. However, with several simplifying assumptions, especially that $\psi$ is independent of $x$ (justified by $d\ll\xi$ in the thin film limit $d\rightarrow0$), it is possible to obtain\cite{TinkhamBook}
\begin{equation}
\begin{aligned}
-\frac{\text{d}^2\psi}{\text{d}z^2}+\bigg(\frac{2\pi\mu_0H\text{sin}(\theta)}{\Phi_0}\bigg)^2&z^2\psi=\\
\Bigg[\frac{1}{\xi^2}-\bigg(&\frac{\pi d\mu_0H\text{cos}(\theta)}{\sqrt{3}\Phi_0}\bigg)^2\Bigg]\psi
\end{aligned}
\end{equation}
The structure of this equation is completely equivalent to the one-dimensional Schr\"odinger equation of the harmonic oscillator, describing a particle of mass $m$ in a harmonic potential $Dx^2/2$ (with $D$ the spring constant), given by\cite{Gasiorowicz74}
\begin{equation}
\left(-\frac{\hbar^2}{2m}\frac{\text{d}^2}{\text{d}x^2}+\frac{m\omega^2}{2}x^2\right)u(x)=Eu(x)
\end{equation}
with the angular frequency $\omega=\sqrt{D/m}$ and the eigenvalues \mbox{$E=(n+1/2)\hbar\omega$} with \mbox{$n= 0,1,2\ldots$} the quantum number. In this case the eigenvalue $E$ is given by $(n+1/2)$ multiplied by twice the square root of the product of the prefactor of $-\text{d}^2/\text{d}x^2$ and the prefactor of $x^2$, that means $E=(n+1/2)\cdot2\left\{\left[\hbar^2/\left(2m\right)\right]\left[m\omega^2/2\right]\right\}^{1/2}$.\\
Applying this procedure to Eq.(12), identifying $x$ with $z$, yields
\begin{equation}
E=(n+1/2)\cdot2\left[1\cdot\left(\frac{2\pi\mu_0H\text{sin}(\theta)}{\Phi_0}\right)^2\right]^{1/2}
\end{equation}
On the other hand, it is
\begin{equation}
E=\frac{1}{\xi^2}-\bigg(\frac{\pi d\mu_0H\text{cos}(\theta)}{\sqrt{3}\Phi_0}\bigg)^2
\end{equation}
For $n=0$ the magnetic field for a given $\theta$ becomes maximal, so that $H=H_c(\theta)$, yielding
\begin{equation}
\left|\frac{2\pi\xi^2\mu_0}{\Phi_0}H_c(\theta)\text{sin}(\theta)\right|+\bigg(\frac{\pi\xi d\mu_0}{\sqrt{3}\Phi_0}H_c(\theta)\text{cos}(\theta)\bigg)^2=1
\end{equation}
Finally, equating the coefficients in Eq.(16) with Eqs.(3) and (5) results in Tinkham's formula, given in Eq.(6).\\
\subsubsection{Extensions of the Ginzburg-Landau Theory and the Theory of Type II Superconductors}
Extensions of the microscopic version of the GL theory and the theory of type II superconductors in a magnetic field towards lower temperatures were carried out by Maki\cite{Maki64,Maki64-2,Maki66}, Maki and Tsuzuki\cite{Maki65}, de Gennes\cite{deGennes64}, Caroli et al.\cite{Caroli66}, Tewordt\cite{Tewordt63,Tewordt65,Tewordt64,Tewordt65-2}, Neumann and Tewordt\cite{Neumann66,Neumann66-2}, Werthamer\cite{Werthamer63}, Helfland and Werthamer\cite{Helfland64,Helfland66}, Werthamer et al.\cite{Werthamer66}, and Werthamer and McMillan\cite{Werthamer67}. The topic is reviewed by Werthamer\cite{Werthamer69}, Cyrot\cite{Cyrot73}, and Fetter and Hohenberg\cite{Fetter69}. The ranges of validity of the extensions of the GL theory are summarized in Fig. 6 of Werthamer's review\cite{Werthamer69}. The ranges of applicability of the extensions to the description of type II superconductors are given in Fig. 13 of the Fetter and Hohenberg review\cite{Fetter69}. There is a range of applicability in a certain region of magnetic fields close to the $H_{c2}(T)$ line down to zero temperature. Nevertheless, there is no application of the results to get a 'Tinkham-like' formula for an extended temperature range (as far as known to the authors). The reason may be the complexity of the theoretical expressions, which often only allow a numerical solution.\\
\subsubsection{Niobium - An Intermediate Coupling Superconductor}
In the present work, Nb is used as S-material. According to Finnemore et al.\cite{Finnemore66} Nb is not a weak coupling, but an intermediate coupling superconductor. A quantity, characterizing the strength of the electron-phonon coupling, is the parameter $\lambda$, describing the effective mass enhancement, $m^*/m$, from the effective mass $m$ of the electron determined by the band structure, due to the electron-phonon interaction, given by\cite{McMillan68,Scalapino69} $m^*/m=1+\lambda$. The value of $\lambda$ for Nb is determined to be in the range of\cite{McMillan68,Weber91,Carbotte90} $0.8$ to $1.2$, which is between the values\cite{Carbotte90,Bergmann73} for In ($\lambda=0.8$) and Hg  ($\lambda=1.6$). Indium can be regarded as almost weak coupling superconductor, while mercury is a strong coupling superconductor.\\ Another measure for the strength of the electron-phonon interaction is the ratio $2\mathbf{\Delta}(0)/(kT_{c0})$, where $\mathbf{\Delta}(0)$ is the energy gap at zero temperature. The prediction of the BCS theory, valid for weak coupling superconductors, of this ratio is\cite{TinkhamBook} $3.5$. For Nb values between $3.6$ and $3.8$ are obtained from experiments\cite{Buckel94,Gladstone69}, which are more close to the BCS value obtained for Sn and In, than to $4.3$ and $4.6$ obtained for the strong coupling superconductor Pb and Hg, respectively, obtained from tunneling experiments\cite{Buckel94}.\\
Werthamer and McMillan\cite{Werthamer67} calculated the strong coupling corrections to $H_{c2}(T)$ and carried out a numerical computation for Nb. They found, that the strong coupling effects constitute only a negligible portion of the discrepancy between the weak coupling theory and the experimental observation, which is mainly caused by Fermi surface anisotropy.\\
Thus, we will apply the weak coupling results of Chap.~III of the main text to the data for Nb films and the S/F bilayer of the present work.\\
\subsubsection{Validity Conditions for the Tinkham Formula}
With the parameters obtained in the main text, we now investigate, whether the conditions for the validity of Tinkham's formula are fulfilled, \textit{i.e.} if $d<\sqrt{5}\lambda(T)$ and $d<<\xi(T)$. Since studies of $\lambda(T)$ and $\xi(T)$ are not available for S/F bilayers, this can strictly only be done for Nb5/1. However, we will test the conditions also for S23\#5 under different assumptions.\\
The magnetic penetration depth of thin superconducting Nb films has been measured\cite{Siegel,Lemberger}. For film thicknesses between $7$~nm and $20$~nm, $\lambda(T=0~$K$)$ decreases from about $240$~nm to $140$~nm (see Fig. 6 of the work of Gubin et al.\cite{Siegel}). Since $\lambda(T)$ increases for increasing temperature, the values of $\lambda(T)$ are expected to be always much larger than the film thickness of $7.3$~nm in Nb5/1 and a Nb layer with a thickness of $14.1$~nm, as present in S23\#5. We expect, that $\lambda(0)=0.5^{1/2}\chi^{-1/2}\lambda_L(0)$ is increasing for a Nb film in the presence of an F-layer ($\lambda_L(0)$ should not change, $\chi$ should decrease, because $\xi_0$ increases, for details see below). From the phenomenological GL theory one gets\cite{Tidecks90} $\lambda(T)=(m'/(e'^2\mu_0\left|\psi_0\right|^2)^{1/2}$, where $\left|\psi_0\right|^2=n_{s0}$. Since the pairing wave function in the F-layer is smaller than in the S-layer, \textit{i.e.} the superconducting charge carrier density is reduced, one expects (identifying $|\psi_0|^2$ with $|\Phi_{F}|^2$) an even larger magnetic penetration depth there. Thus, in the investigated samples the condition $d<\sqrt{5}\lambda(T)$ is expected to be fulfilled for all temperatures.\\
Next, we calculate the GL coherence length for Nb5/1 and a freestanding Nb-film of thickness $14.0$~nm (similar to the one present in S23\#5). To calculate $\xi(T)$, we use that $\xi(0)$ is $9.7$~nm and $10.5$~nm for Nb5/1 and the Nb-film of thickness $14.0$~nm, respectively (see Chap.~IV-A of the main text). The critical temperatures of the films are $5.95$~K and $8.00$~K, respectively. Using $\xi(T)=\xi(0)(T_{c0}/(T_{c0}-T))^{1/2}$, we obtain $108$~nm and $22$~nm at $T=T_{MP}=5.90$~K and 6.10~K for Nb5/1 and S23\#5, respectively. Thus, $d=d_S<<\xi(T)$ is fulfilled for Nb5/1 and at least $d_S<\xi(T)$ is fulfilled for a freestanding Nb-film with similar $d_S$ as the one in S23\#5.\\
For the further discussion, we now calculate BCS coherence lengths according to the expression given in Chap.~III of the main text. Using $v_F=2.768\cdot10^5$~m/s for Nb, according to Weber et al.\cite{Weber91}, and inserting the respective critical temperatures yields $\xi_0=64.0$~nm and $47.7$~nm for Nb5/1 and the freestanding Nb film of $14.0$~nm, respectively. With $\xi(0)=0.74\chi^{1/2}\xi_0$ (see Eq.(7)), we then get $l=2.1$~nm and $3.5$~nm, respectively. Thus, both Nb films are in the dirty limit ($l<<\xi_0$).\\
To estimate $\xi(T)$ for the Nb film in S23\#5, we assume that its critical temperature is only suppressed due to the proximity effect by the F-layer. Thus, the $14.0$~nm freestanding Nb film, however, with a suppressed $T_{c0}$ of $6.34$~K, is a suitable reference system. For this (fictive) film, we obtain, according to Eq.(7), $\xi(0)=11.9$~nm, using $\xi_0=60.2$~nm and $l=3.5$~nm. Consequently, we obtain $\xi(T_{MP}=6.10$~K$)=61.3$~nm and, thus, again it is $d=d_S<<\xi(T)$.\\
To get an estimate of $\xi(T)$ for the whole sample S23\#5, we use $\xi(0)=12.8$~nm, as obtained in \mbox{Chap.~IV-A} of the main text, yielding $\xi(T_{MP}=6.10$~K$)=65.9$~nm, so that $d=d_S+d_F<\xi(T)$.\\
Here, we do not consider an enhancement factor to the slope $d\mu_0H_{c\perp}/dT$, calculated for Nb by Butler \cite{Butler80} (see the discussion by Weber et al. \cite{Weber91}, in our case the factor for the dirty limit \cite{Butler80} would be appropiate). This would lead to a slightly larger value of $\xi(0)$ and, thus, $l$. However, this would not change the presented conclusions.\\
The expressions for the magnetic penetration depth and the GL coherence entering the derivation of the Tinkham formula are those in a weak magnetic field\cite{Tinkham, TinkhamBook}, given in Chap. III of the main text. Thus, it is not necessary to consider a possible magnetic field dependence of these quantities. According to Douglass\cite{Douglass61}, for thin superconducting films in a parallel magnetic field, $\lambda(T,H)$ is larger than $\lambda(T,0)$ and approaches infinity for $H\rightarrow H_c$. Kogan proposed a magnetic field dependence of the coherence length $\xi(T,H)<\xi(T,0)$\cite{Kogan86}, which can be neglected in the dirty case, and a resulting influence on the superconducting transition temperature \cite{Kogan86,Kogan87}. However, this theory is controversially discussed in the literature\cite{Scotto91,Hara94,Seguchi92}. According to Kogan\cite{Kogan86} the proposed enhancement of the transition temperature should occur for $d_S$ below a critical thickness, $d_c$. For Nb5/1 it is, however, $d_S>d_c=5.7$~nm. For S23\#5, there is no uniform $l$, so $d_c$ cannot be calculated. In any case, we do not observe any evidence for this enhancement in both samples.\\
\subsection{Oscillation Properties of the Pairing Wave Function in the Quasi-One-Dimensional FFLO-Like State in S/F Bilayers}
The oscilatory behavior of $\Phi_F$ has been analyzed theoretically in detail\cite{Tagirov98}. The oscillation wavelengths and decay lengths for the case of a clean and dirty ferromagnet are summarized in Chap. IV of our previous work\cite{Sidorenko10}. Moreover, the topic is discussed in detail in the Appendix of the doctoral thesis of Kehrle\cite{Kehrle12}, where it is also shown, that the decay length in the clean case is given by twice the electron mean free path, $l_F$ in the F-material (the factor of 2 was omitted in our previous work \cite{Sidorenko10}).\\
For $\Phi_F(x_F)\propto \text{cos}(k_{FM}x_F)$ the pairing wave function has its first node at $k_{FM}x_{F1}=\pi/2$ and its first minimum at $k_{FM}x_{F2}=\pi$. Here, $k_{FM}=2\pi/\lambda_{FM}$ is the wave number and $\lambda_{FM}$ the oscillation wavelength. We thus get $x_{F1}=\lambda_{FM}/4$ and $x_{F2}=\lambda_{FM}/2$. Since the experimental results for oscillatory behavior of $T_c(d_F)$ are best described by the extension of the dirty case theory towards the clean case, as discussed in Chap. IV of our previous work\cite{Sidorenko10}, we apply the clean case expression for the oscillation wavelength, \textit{i.e.} $\lambda_{FM}=\lambda_{F0}=2\pi\xi_{F0}$. where $\xi_{F0}=\hbar v_F/E_{ex}$, with $E_{ex}$ the exchange splitting energy. According to our previous work\cite{Sidorenko10}, for sample series S23, it is $\xi_{F0}=10.8$~nm, yielding $\lambda_{FM}=67.9$~nm and, thus, $x_{F1}=17.0$~nm and $x_{F2}=33.9$~nm.\\
As discussed in our previous work\cite{Sidorenko09}, for $d_F=x_{F1}$, the reflection of $\Phi_F$ at the outer border of the F-layer leads to interference effects yielding the first minimum of $T_c(d_F)$. The experimental results for $T_c(d_F)$ of sample series S23 are shown in Fig. 6 of our previous work\cite{Sidorenko10}, yielding this minimum at $d_F=7.0$~nm. The experiments are well described by the theory. Thus, there is a phase shift of the pairing wave function at the S/F interface due to boundary conditions, so that $\Phi_F(x_F)\rightarrow \Phi_F(x_F+10$~nm~$)$ and, thus, $x_{F1}\rightarrow 7.0$~nm and $x_{F2}\rightarrow 23.9$~nm. Consequently, the distance of the first minimum of $\Phi_F(x_F)$ to the S/F interface is 23.9~nm.\\
\bibliography{final}


%
%

%



\end{document}